\documentclass[11pt, a4paper]{article}
\usepackage{jheparxiv}
\usepackage[utf8]{inputenc}
\usepackage{amsmath}
\usepackage{amsfonts}
\usepackage{amssymb}
\usepackage{latexsym}
\usepackage{mathrsfs}
\usepackage{braket}		
\usepackage{graphicx}
\usepackage{amsmath}
\usepackage{amsfonts}
\usepackage{amssymb}
\usepackage{latexsym}
\usepackage{mathrsfs}
\usepackage{color}
\usepackage{xcolor}
\usepackage{slashed}
\usepackage{twistor}
\usepackage{tikz-cd}
\usepackage{mathtools}
\usepackage{bm}
\usepackage{color}
\usepackage{soul}
\usepackage{braket}
\usepackage{float}
\usepackage{comment}
\usepackage[]{todonotes}
\usepackage{color}
\usepackage{xcolor}
\usepackage{slashed}
\usepackage{twistor}
\usepackage{tikz}
\usepackage{subcaption}
\usepackage{mathtools}
\usepackage{bbm}
\usepackage{tcolorbox}
\usepackage{feynmp}
\usepackage{epsfig}
\usepackage[dvipsnames]{xcolor}
\def\bd {\boldsymbol}

\renewcommand{\bd}[1]{\begin{fmffile}{#1}\begin{fmfgraph*}}
\newcommand{\ed}{\end{fmfgraph*}\end{fmffile}}

\newcommand{\mrin}{\mathrm{in}}
\newcommand{\mrout}{\mathrm{out}}

\renewcommand{\d}{\mathrm{d}}

\graphicspath{{./figures/}}

\title{Leading soft theorems on plane wave backgrounds}
\author{Sonja Klisch}

\affiliation{School of Mathematics and Maxwell Institute for Mathematical Sciences \\
        University of Edinburgh, EH9 3FD, United Kingdom}

\emailAdd{s.klisch@ed.ac.uk}

\abstract{The infrared singularities of scattering amplitudes have historically contributed to much development in understanding fundamental structures in physics. However, the fate of the leading soft singularities of amplitudes in non-trivial background fields has remained largely unknown. In this paper, we derive the leading soft theorems for photons, gluons and gravitons on generic plane wave backgrounds in gauge theory and gravity. The results differ from the flat space results through dependence on the initial conditions of the soft mediator. We also consider the special case of self-dual plane wave backgrounds, and match onto the flat space results when the background is treated perturbatively.}
\begin{document}

\maketitle
\section{Introduction}

Soft theorems are a universal feature of scattering amplitudes in theories with massless force mediating particles, such as photons, gluons or gravitons~\cite{Bloch:1937pw,Low:1954kd,Gell-Mann:1954wra,Low:1958sn,Yennie:1961ad,Weinberg:1964ew,Weinberg:1965nx}. They describe the soft singularity when one such external massless mediator is taken `soft' --- approaching zero energy and momentum. In flat space, these soft contributions factorise and, in electromagnetism and gravity, exponentiate~\cite{Bloch:1937pw,Weinberg:1965nx}.  The soft divergence then cancels against infrared singularities arising from virtual photons or gravitons in loops. This cancellation also appears naturally when the external states are dressed appropriately with soft particles~\cite{Kulish:1970ut}. 

Even though they can generally be cancelled, these soft divergences themselves also tell us something fundamental about scattering and the spacetime we are scattering in. 
The soft photon and soft graviton theorems are very closely related to the conservation of charge and energy in scattering processes~\cite{Weinberg:1964ew}. Viewing tree-level amplitudes as rational functions of their kinematics, all-multiplicity formulae for gluon and graviton scattering can be deduced from their soft and collinear limits~\cite{Bern:1998sv,Nguyen:2009jk,Hodges:2011wm,Hodges:2012ym}. And the last decade has seen a slew of development arising from viewing soft theorems as the Ward identities of large asymptotic symmetries in gauge theory and asymptotically flat spacetimes~\cite{He:2014cra,He:2015zea,He:2014laa,Campiglia:2015qka,Kapec:2015ena,Strominger:2017zoo,Miller:2021hty}. 

The above story has been restricted to the scattering of particles in flat spacetimes. However, that is not the only space where scattering amplitudes exist and where one may deduce interesting physics from their infrared singularities. Indeed, there has been much recent work on scattering amplitudes in plane wave backgrounds: a class of solutions in gauge theory and gravity with a large number of symmetries~\cite{Bondi:1958aj,Einstein:1937qu,Brinkmann:1925fr,Blau}. Here, the background is treated exactly, and scattering amplitudes are calculated for perturbations satisfying the free equations of motion on this classical solution~\cite{Arefeva:1974jv,Jevicki:1987ax,Adamo:2017nia}. 

These backgrounds are experimentally relevant since they serve as one of the models for high-intensity lasers~\cite{Fedotov:2022ely}. Scattering amplitudes on an electromagnetic plane wave background therefore captures phenomena such as non-linear Compton and Breit-Wheeler scattering that can be observed experimentally. In gravity, a gravitational plane wave captures many of the non-linear intricacies of general relativity~\cite{Penrose:1965rx,Harte:2013dba,Harte:2015ila} and the behaviours of particle scattering on curved spacetimes~\cite{Gibbons:1975jb,Ward:1987ws,Friedlander:2010eqa,Garriga:1990dp,Zhang:2017rno,Zhang:2018srn,Harte:2024mwj}. Gravitational plane waves can also be related to any null geodesic on any spacetime via the Penrose limit~\cite{Penrose:1976,Blau}. However, a limiting factor is the complexity involved in calculating amplitudes in these contexts --- only low numbers of external legs and loops have been calculated for an arbitrary plane wave profile. 

A considerable advantage is gained when the background is self-dual. This is due to the integrability of the self-dual sectors of both Yang-Mills and Einstein gravity. Exploiting the twistor construction for these theories~\cite{Penrose:1976js,Ward:1977ta}, it's possible to obtain all-multiplicity expressions for the maximally helicity violating (MHV) amplitudes for gluon and graviton scattering in these theories on a variety of self-dual backgrounds~\cite{Adamo:2020syc,Adamo:2020yzi,Adamo:2022mev,Adamo:2024xpc,Adamo:2023fbj} including self-dual plane waves. The twistor construction also gives well-motivated formulae for N$^k$MHV scattering amplitudes on these backgrounds, similar to the flat space versions~\cite{Witten:2003nn,Roiban:2004yf,Cachazo:2012kg}. 

Whilst scattering on self-dual backgrounds is also physically relevant --- for example in modelling complete depletion in lasers~\cite{Adamo:2025vzv} --- one may wonder if it is possible to derive all-multiplicity statements in generic non-chiral backgrounds too. An accessible first step in this venture is studying the infrared singularities of amplitudes on a generic plane wave background. Beyond just being an excellent guide for what generic amplitudes may look like, they may shed a light on non-perturbative aspects of celestial holography and new approaches to perturbing around the self-dual sector. Previous work on generic plane wave backgrounds has included treatment of infrared divergences from soft photons at all multiplicities~\cite{Ilderton:2012qe,Dinu:2012tj}, which we refer to when we consider photons.

\medskip

In this paper, we study the leading soft singularities of scattering amplitudes in plane wave backgrounds in gauge theory and Einstein gravity. We restrict to the soft singularities in the scattering of minimally coupled (massive or charged) scalars in gauge theory and gravity, with one soft mediator. Whilst not explored explicitly in this paper, coupling to other particle types is expected to proceed analogously. We find that the soft singularities that arise generally depend on the asymptotics of the soft particle --- whether it is inserted in the ingoing, or the outgoing region. A property of solving the wave equation on these backgrounds is that generally a soft ingoing particle will not look the same in the outgoing region, and this is reflected in the soft factorisation. 

The results of this paper are:
\begin{tcolorbox}[colback=red!5!white,colframe=red!75!black,title=Soft photon theorem on a gauge theory plane wave background~\cite{Ilderton:2012qe,Dinu:2012tj}]\vspace{-0.35cm}
    \begin{equation}
        \lim_{\omega \rightarrow 0} \omega \mathcal{M}_{n + 1}(\{p_i\}; \omega \hat{k}) = e \,\Bigg[ \sum_{p_i \text{ ingoing}} Q_i\,\frac{\epsilon \cdot p_i}{\hat{k} \cdot p_i}  + \sum_{p_i \text{ outgoing}} Q_i \,\frac{\epsilon \cdot p_i}{ \hat{k} \cdot p_i} \Bigg] \mathcal{M}_n(\{p_i\})
    \end{equation}
\end{tcolorbox}
\begin{tcolorbox}[colback=red!5!white,colframe=red!75!black,title=Soft gluon theorem on a gauge theory plane wave background]\vspace{-0.35cm}
    \begin{align}
        \lim_{\omega \rightarrow 0} \omega \mathcal{M}_{n+1}^\mathsf{a} (\{ p_i\}; \omega \hat{k}^\mrin) &= g_{\text{YM}}\sum_{p_i \text{ ingoing}} \frac{\epsilon \cdot p_i}{\hat{k} \cdot p_i} \, T^{\mathsf{a}}_{p_i} \, \mathcal{M}_n(\{p_i\}), \\
        \lim_{\omega \rightarrow 0} \omega \mathcal{M}_{n+1}^\mathsf{a} (\{ p\}; \omega \hat{k}^\mrout) &= g_{\text{YM}} \sum_{p_i \text{ outgoing}} \frac{\epsilon \cdot p_i}{\hat{k} \cdot p_i}\, (T^{\mathsf{a}}_{p_i})^{\ast} \,\mathcal{M}_n(\{p\})
    \end{align}
    Here $T^{\mathsf{a}}_{p_i}$ is the Lie algebra generator corresponding to the soft gluon in the representation carried by particle with momentum $p_i$. 
\end{tcolorbox}
\begin{tcolorbox}[colback=red!5!white,colframe=red!75!black,title=Soft photon theorem on a gravitational plane wave background]\vspace{-0.35cm}
    \begin{gather}
        \lim_{\omega \rightarrow 0} \omega \mathcal{M}_{n+1}(\{p_i\}; \omega \hat{k}^\mrin) = e \sum_{p_i \text{ ingoing}} Q_i \, \frac{\epsilon \cdot p_i}{\hat{k} \cdot p_i}\, \mathcal{M}_n(\{p_i\}), \\
        \lim_{\omega \rightarrow 0} \omega \mathcal{M}_{n+1} (\{p\}, \omega \hat{k}^\mrout) =  e\sum_{p_i \text{ outgoing}} Q_i \, \frac{\epsilon \cdot p_i}{\hat{k} \cdot p_i} \, \mathcal{M}_n(\{p_i\})
    \end{gather}
\end{tcolorbox}
\begin{tcolorbox}[colback=red!5!white,colframe=red!75!black,title=Soft graviton theorem on a gravitational plane wave background]\vspace{-0.35cm}
    \begin{multline}
    \lim_{\omega \rightarrow 0} \omega \mathcal{M}_{n+1}(\{p_i\}; \omega \hat{k}^\mrin)  \\  = \frac{\kappa}{2} \sum_{p_i \text{ ingoing}} \, \Bigg[ \frac{(\epsilon \cdot p_i)^2}{\hat{k} \cdot p_i}-   \frac{p_{i \, +}}{\hat{k}_+} \, \epsilon^a \epsilon^b \,  \int_{-\infty}^{\infty}\frac{\sigma^\mrin_{ab}(s)}{|E^{\mrin}(s)|^{1/2}} \d s\Bigg]\, \mathcal{M}_n(\{p_i\}), \label{gravitonGR1}\end{multline}
    \begin{multline}\lim_{\omega \rightarrow 0} \omega \mathcal{M}_{n+1} (\{p\}, \omega \hat{k}^\mrout) \\= \frac{\kappa}{2} \sum_{p_i \text{ outgoing}}  \,\Bigg[ \frac{(\epsilon \cdot p_i)^2}{\hat{k} \cdot p_i} -  \frac{p_{i \, +}}{\hat{k}_+} \, \epsilon^a \epsilon^b \,  \int_{-\infty}^{\infty}\frac{\sigma^\mrout_{ab}(s)}{|E^{\mrout}(s)|^{1/2}} \d s\Bigg]\, \mathcal{M}_n(\{p_i\})\label{gravitonGR2}
\end{multline}
where $E_{i\, a}^{\mrin/\mrout}$ and $\sigma_{ab}^{\mrin/\mrout}$ are background dependent functions defined in Section \ref{gravSec}.
\end{tcolorbox}

The above assume a general, strong plane wave background with non-trivial memory. The momenta point into the scattering process, and ingoing/outgoing is determined by the sign of $p_{i \, +}$ in lightfront coordinates. When relevant, the charge of particle $i$  is $Q_i$. The soft particle is treated as propagating on the background, and ingoing and outgoing wavefunctions will look different if it interacts with the backgrond. Here, these labels are determined by whether it looks like a usual flat space soft particle in the ingoing or outgoing regions of the spacetime. Special cases for self-dual and weak backgrounds are explored in Section \ref{sec:Comp}, with soft theorems presented when factorisation at tree level can be shown.

\medskip

This paper is organised as follows. In Section \ref{sec:bg} we review plane wave backgrounds in gauge theory and in gravity. The special case of self-dual backgrounds is also discussed. In Section \ref{sec:soft} we calculate the leading soft singularities arising from soft photons, soft gluons and soft gravitons in these backgrounds. We also comment on the gauge invariance of the results. Consequently, in Section \ref{sec:Comp} we contrast with results in self-dual backgrounds (taking the MHV amplitude as a comparison) and derive the perturbative and flat space results as special cases. Finally, we conclude in Section \ref{sec:Conc}. Appendix \ref{app:Feyn} collects some basic Feynman rules for calculating scattering amplitudes on plane wave backgrounds, whilst Appendix \ref{sec:conv} presents the convergence properties of some integrals encountered in the text.

\medskip

Throughout this paper we consider scattering in four dimensions and in light-front coordinates $(x^+, x^-, x^{\perp})$ where 
\begin{equation}
x^+ = \frac{x^0 + x^3}{\sqrt{2}}, \qquad x^- = \frac{x^0 - x^3}{\sqrt{2}},
\end{equation}
and $x^{\perp}$ is shorthand for the transverse Cartesian coordinates $x^a$ with $a = 1, 2$. In these coordinates the flat Minkowski metric in mostly minus signature is
\begin{equation}
    \d s^2_{\text{flat}} = 2 \d x^+ \d x^- - \delta^{ab} \d x^a \d x^b 
\end{equation}
We will also use $\hat{\d} = \d / 2\pi$ throughout.

\paragraph{Erratum (v2):} In the first version of this paper, it was incorrectly claimed that no soft graviton theorem holds on generic gravitational plane waves. After reviewing the calculation in Section \ref{sec2:grav} in the first version, a mistake was found in treatment of the stationary phase approximation and the $\omega \rightarrow 0$ limit. In this version, we have addressed this error, and revised the claims of the paper accordingly. The results for gluons and photons remain unaltered.

\section{Plane wave backgrounds in gauge theory and gravity} 
\label{sec:bg}

Our backgrounds of interest in both gravity and gauge theory are plane wave backgrounds. These are backgrounds with a large number of symmetries that allow for exact solutions to the linearised equations of motion of their perturbations. They describe pure radiation of either the Maxwell field or the gravitational field, propagating between past and future null infinity along some given constant null direction, herein labelled by $n^{\mu}$. From a particle physics perspective, they can also be interpreted as a coherent superposition of collinear photons or gravitons, and this picture will be useful for interpreting some of the results in this paper. But crucially we will treat these backgrounds as `strong' with no perturbative regime \emph{a priori}, and will comment at several steps where this condition is crucial.

We will only consider sandwich plane waves: those where non-zero field strength or curvature is only compactly supported in a region $[x_i^-, x_f^-]$ in the $x^-$ light-front coordinate. This gives us flat \emph{ingoing} ($x^- < x_i^-$) and \emph{outgoing} ($x^- > x_f^-$) regions.  

\subsection{Gauge theory}

In gauge theory (both electromagnetism and Yang-Mills) we consider a solution to the vacuum equations of motion with gauge potential 
\begin{equation}
    \mathsf{A} = - \mathsf{A}_a (x^-) \,\d x^a \label{sec1:gaugeER}
\end{equation}
valued in the Cartan subalgebra of the gauge group, $\mathfrak{h} \subset \mathfrak{g}$. It is possible to transform these to a less ambiguous gauge via the gauge transformation $\mathsf{A} \rightarrow \mathsf{A} + \d (x^a \mathsf{A}_a)$ which gives 
\begin{equation}
    \mathsf{A} = x^a \dot{\mathsf{A}}_a  \,\d x^-. \label{sec1:gaugeBr}
\end{equation}
Non-abelian pp-waves satisfying the vacuum Yang-Mills equations of motion have also been considered in~\cite{Coleman:1977ps,Trautman:1980bj} and take the same functional form as \eqref{sec1:gaugeBr}. However they are no longer plane waves due to no longer having the defining Heisenberg symmetry group~\cite{Trautman:1981qm,Adamo:2017nia}.

The field strength of the Cartan-valued potential is 
\begin{equation}
    F = \dot{\mathsf{A}}_a \,  \d x^a \wedge \d x^-,
\end{equation}
which can be easily checked to satisfy the Maxwell equations, and hence Yang-Mills equations for the Cartan subalgebra. Asserting that the plane wave is sandwich means that the field strength is compactly supported on $x_i^- < x^- < x_f^-$. This in particular means that $\dot{\mathsf{A}}_a = 0$ outside of this region. Looking back at \eqref{sec1:gaugeER}, we have two convenient gauge choices: the \emph{ingoing} gauge and the \emph{outgoing} gauge
\begin{equation}
    \mathsf{A}^{\mrin}_a(x^-< x_i^-) = 0, \qquad  \mathsf{A}^{\mrout}_a (x^-> x_f^-) = 0. \label{sec1:gaugeInOut}
\end{equation}
The two are related by a constant $a_{\infty \, a} = \mathsf{A}^{\mrout}_a - \mathsf{A}^{\mrin}_a $ where 
\begin{equation}
    a_{\infty \, a} = \int^{x_f^-}_{x_i^-} \dot{\mathsf{A}}_a \, \d x^-. \label{sec1:gaugeMem}
\end{equation}
This quantity can be viewed as proportional to the `work done' on charged matter passing through the background. It can also be interpreted as encoding the \emph{electromagnetic memory effect} for these backgrounds. 

In this paper we mainly assume that $a_{\infty \, a} \neq 0$, and indeed much larger than the soft parameters we will consider. Therefore we neglect those `miraculous' backgrounds where the ingoing and outgoing gauges \eqref{sec1:gaugeInOut} are equivalent (and hence $a_{\infty \, a} = 0$). What this also means is that we can not treat the background as `weak' as \eqref{sec1:gaugeMem} is proportional to the strength of the background field.  This condition is relaxed in Section \ref{sec:weakYM}.

\medskip

We will now proceed to solve the linearised equations of motion on this background.

\paragraph{Charged scalars.}
The free equation of motion of a massless scalar with charge $e$ with respect to the Cartan-valued background and mass $m$ is 
\begin{equation}
    D_{\mu} D^{\mu} \Phi(x) = \Big(2 \partial_- \partial_+ - \partial_\perp \partial^\perp  - 2 \im e \, x^a  \dot{\mathsf{A}}^a \, \partial_+ \Big) \phi(x) = - m^2 \Phi(x),
\end{equation}
with covariant derivative $D_{\mu} = \partial_{\mu} - \im e \mathsf{A}_{\mu}$ in the gauge \eqref{sec1:gaugeBr}\footnote{Working in this gauge ensures that both the ingoing and outgoing regions have zero potential and we can define compatible ingoing and outgoing states.}. This can be solved~\cite{Wolkow:1935zz} by $\Phi(x) = e^{\im \phi_k(x)}$ with 
\begin{multline}
    \phi_k(x) =    k_+ x^+ + (k_\perp + e \mathsf{A}_{\perp}(x^-) ) x^{\perp} + \frac{1}{2k_+} \int^{x^-} \d s \, (k_\perp + e \mathsf{A}_{\perp} (s))(k^\perp + e \mathsf{A}^{\perp} (s))\\ + \frac{m^2}{2k_+}\, x^- \label{sec1:phisol}
\end{multline}
where $(k_+, k_\perp)$ are arbitrary, specifying the degrees of freedom of the on-shell momentum. Note that this solution also gives some intuition for the geodesic motion of the associated particle in this background. We can define the dressed momentum $K_{\mu} = e^{- \im \phi_k} D_{\mu} e^{\im \phi_k}$ having the form
\begin{equation}
    K_{\mu} \d x^{\mu} = k_+ \d x^+ + (k_a + e \mathsf{A}_a) \, \d x^a + \frac{1}{2k_+} \Bigg((k_a + e \mathsf{A}_a) (k^a + e \mathsf{A}^a) + m^2 \Bigg) \, \d x^-. \label{sec1:dMom}
\end{equation}
After passing through the wave, a particle of charge $e$ will gain transverse momentum $e a_{\infty }$ whilst remaining on-shell.

The scalar solution depends on the boundary conditions we impose on it. The natural boundary conditions to consider for a scattering problem are 
\begin{equation}
    \Phi^{\mrin}_k(x^- < x_i^-) = e^{\im k \cdot x}, \qquad \Phi^{\mrout}_k(x^-> x_f^-) = e^{\im k \cdot x}.
\end{equation}
These precisely correspond to choosing a gauge, as in \eqref{sec1:gaugeInOut}, for $\mathsf{A}_a$ appearing in \eqref{sec1:phisol}. The ingoing and outgoing solutions are then simply 
\begin{equation}
    \Phi^{\mrin}(x) = e^{\im \phi_k^{\mrin}(x)}, \qquad \Phi^{\mrout}(x) = e^{\im \phi_k^{\mrout}(x)}
\end{equation}
where the superscripts are understood to modify $\mathsf{A}_a \rightarrow \mathsf{A}^{\mrin/\mrout}_a$ in the phase.

\paragraph{Photons.} Photons (equivalently viewed as Cartan-valued perturbations) do not `see' the background, so the solution to the equation of motion is the flat one
\begin{equation}
    a_{\mu}(x) = \epsilon_{\mu} e^{\im k \cdot x}.
\end{equation}
This solution has the same functional form for both ingoing and outgoing boundary conditions. Here (and henceforth) we will choose polarisations to satisfy light-front and Lorenz gauge: $n^{\mu} \epsilon_{\mu} = k^{\mu} \epsilon_{\mu} = 0$.

\paragraph{Gluons.} In contrast, gluons, valued outside of the Cartan subalgebra of the colour algebra satisfy the background-dependent equations of motion
\begin{equation}
    D_{\mu} (D^{\mu} a^{\nu} - D^{\nu} a^{\mu}) + a_{\mu} (\partial^{\mu} \mathsf{A}^{\nu} - \partial^{\nu} \mathsf{A}^{\mu}) = 0.
\end{equation} 
The solution in lightfront and Lorenz gauge can be constructed from the scalar equation~\cite{MasonSpin} and takes the form 
\begin{equation}
    a_{\mu} = \mathcal{E}_{\mu}(x^-) \, e^{\im \phi_k},  \quad \mathcal{E}_{\mu}(x^-) \coloneqq \epsilon_a \delta^a_{\mu} + \frac{1}{k_+} (k^a + e \mathsf{A}^a)\epsilon_a n_{\mu}
\end{equation}
with the massless form of $\phi_k$ which we recall for convenience
\begin{equation}
    \phi_k(x) =    k_+ x^+ +(k_\perp + e \mathsf{A}_{\perp}(x^-)) x^{\perp} + \frac{1}{2k_+} \int^{x^-} \d s \, (k_\perp + e \mathsf{A}_{\perp} (s))(k^\perp + e \mathsf{A}^{\perp} (s)) \label{sec1:massless}
\end{equation}. The polarisation is specified entirely by the transverse components $\epsilon_a$, $a \in \{1, 2\}$, matching the physical degrees of freedom of spin-1 perturbations in four dimensions. The charge $e$ is the charge of the adjoint-valued gluon with respect to the background generator. The dressed polarisation $\mathcal{E}_{\mu}$ satisfies the `dressed' versions of the gauge conditions: $n^{\mu} \mathcal{E}_{\mu} = K^{\mu} \mathcal{E}_{\mu} = 0$. 

Like the scalar case, one must specify boundary conditions on these fields to define the scattering problem. We again want  
\begin{equation}
    a^{\mrin}_{\mu} (x^- < x_i^-) = \epsilon_{\mu} e^{\im k \cdot x}, \qquad a^{\mrout}_{\mu} (x^-> x_f^-) = \epsilon_{\mu} e^{\im k \cdot x}.
\end{equation}
As before, this just corresponds to $\mrin/\mrout$ labels on $\mathsf{A}^a$, including on the dressed polarisation, and so we have
\begin{equation}
    a^{\mrin}_{\mu}(x) = \mathcal{E}_{\mu}^{\mrin}(x^-) e^{\im \phi^{\mrin}_k}, \qquad a^{\mrout}_{\mu}(x) = \mathcal{E}_{\mu}^{\mrout} (x^-) e^{\im \phi^{\mrout}_k}.
\end{equation}

\subsubsection{Self-dual plane waves in gauge theory}
It will be interesting to contrast some results of the next section to the case where we consider a plane wave that is \emph{self-dual}, in the sense that the field strength satisfies $ \star F = i F $. Self-dual gauge theory plane waves are described by a single function $f(x^-)$ and have potential 
\begin{equation}
    \mathsf{A}^+_a = \begin{pmatrix}
        - f(x^-) \\ \im f(x^-)
    \end{pmatrix}.
\end{equation}
Importantly, this is a complex-valued solution. Equivalently,  one may parametrise the transverse coordinates by $z = x_1 + i x_2, \bar{z} = x_1 - i x_2$ in which case we have $\mathsf{A}^+_a \d x^a = - f(x^-) \,\d \bar{z}$. The key properties of this field for our purposes is that 
\begin{equation}
    \mathsf{A}^+_a \mathsf{A}^{+ \, a} = 0 \quad \text{and} \quad \mathsf{A}^+_a \epsilon^{+ \, a} = 0,
\end{equation}
where $\epsilon^+_a$ is the transverse polarisation vector of a positive helicity photon or gluon. Remarkably, self-duality regularises much of the divergent behaviour of solutions to the free field equations in plane waves. The massless scalar solution \eqref{sec1:massless} now has phase
\begin{equation}
    \phi_k(x) = k_+ x^+ + (k_\perp + e \mathsf{A}_{\perp}(x^-)) x^\perp + \frac{k_\perp k^\perp}{2 k_+} x^- + \frac{1}{k_+} \int^{x^-} k_{\perp} \mathsf{A}^{+ \, \perp}(s) \, \d s 
\end{equation} 
which stays finite as $k$ becomes soft in a non-chiral way (in contrast to rapidly oscillating in the generic case).  The dressed polarisation for negative helicity gluons also stays regular as they become soft:
\begin{equation}
    \mathcal{E}_{\mu}^{(-)}(x^-) = \epsilon_a^{(-)} \delta^a_{\mu} + \frac{1}{k_+} k^a \epsilon_a^{(-)} n_{\mu}.
\end{equation}
On the other hand, positive helicity particles still have a divergent polarisation. These are equivalent to the free fields derived from the twistor construction on self-dual plane waves~\cite{Adamo:2020yzi}.

\subsection{Gravity} \label{gravSec}

Non-linear gravitational plane waves are commonly described in terms of two standard coordinate systems. The one manifesting most of the symmetries are known as Einstein-Rosen coordinates~\cite{Einstein:1937qu} in which the metric is 
\begin{equation}
    \d s^2 = 2 \, d X^- \d X^+ - \gamma_{ij}(X^-) \,\d X^i \d X^j, \label{sec1:ERmetric}
\end{equation}
where $i, j, \ldots = 1, 2$ label the transverse coordinates. Here we see that the translation symmetry in $X^+, X^i$ are preserved, corresponding to the Killing vectors $\frac{\partial}{\partial X^+}, \frac{\partial}{\partial X^i}$. Compared to flat space, we also retain a class of Killing vectors combining boosts and rotations 
\begin{equation}
    \mathcal{X}^i = X^i \,\frac{\partial}{\partial X^+} + F^{ij}(X^-)\, \frac{\partial}{\partial X^{j}}, \quad F^{ij}(X^-) \coloneqq \int^{X^-} \d s \, \gamma^{ij}(s).
\end{equation}
These together form the five generators of a Heisenberg algebra. Whilst Einstein-Rosen coordinates are suited for understanding the symmetries, they will generally have singularities corresponding to the focussing of geodesic congruences in these spacetimes~\cite{Penrose:1965rx,Bondi:1989vm}. The global coordinates we will primarily use to describe gravitational plane waves are Brinkmann coordinates~\cite{Brinkmann:1925fr} 
\begin{equation}
    \d s^2 = 2 \, \d x^- \d x^+ - H_{ab}(x^-)\,  x^a x^b \, (\d x^-)^2 - \d x_a \d x^a.
\end{equation}
The wave profile $H_{ab}(x^-)$ can be a free function (with compact support in our context), with the only condition from the vacuum Einstein equations imposing that it is tracefree: $H^a_a = 0$. For completeness, the non-vanishing components of the curvature (up to index permuations) are 
\begin{equation}
    R^a_{- b -} = - H^a_b (x^-).
\end{equation}

Brinkmann coordinates are related to Einstein-Rosen coordinates via the coordinate transformation
\begin{align}
    X^- &= x^-, \\
    X^+ & = x^+ + \frac{1}{2} \dot{E}^i_a(x^-) \, E_{b \, i}(x^-) \,  x^a x^b , \\
    X^i &= E^i_a(x^-) \, x^a. 
\end{align}
Note that Roman letters early in the alphabet $a, b, \ldots$ will be used to describe objects naturally associated with Brinkmann coordinates, whilst letters in the middle $i, j, \ldots$ describe objects associated with Einstein-Rosen coordinates. Here the transverse vielbein $E_{i \, a}(x^-)$ satisfies the second-order differential equation 
\begin{equation}
    \ddot{E}_{i \, a} = H_{ab} E^{b}_i \label{sec1:VielEq}
\end{equation}
and defines an Einstein-Rosen transverse metric representation for the profile $H_{ab}$ 
\begin{equation}
    \gamma_{ij} = E^a_{(i} E_{j) \, a}. \label{sec1:ERmet}
\end{equation}
The Brinkmann indices on the vielbein are raised and lowered by $\delta_{ab}$, for example $E_{i \, a} \delta^{ab} = E_i^a$. Its inverse is denoted with raised Einstein-Rosen index $E^i_a$, satisfying $E_{i \, a} E^i_b = \delta_{ab}$. 

A useful quantity describing the evolution of geodesic congruences is the deformation tensor 
\begin{equation}
    \sigma_{ab} = \dot{E}^i_a E_{b \, i}, \label{sec1:deform}
\end{equation}
describing the expansion and shear of the $\partial_{X^-}$ null congruence via its trace and trace-free parts respectively. Any quantities built out of the vielbeins and their inverses will have singularities due to the non-globality of the Einstein-Rosen coordinates. These will usually occur as simple poles or zeroes, or, for $\det (E)^{1/2}$, as a branch cut. In the rest of the paper we will be fairly agnostic about these, but it would be interesting to study how to navigate these carefully.

\medskip 

The differential equation \eqref{sec1:VielEq} requires boundary conditions to define a unique solution. This ambiguity means that a plane wave spacetime uniquely described by the sandwich Brinkmann metric can be represented by many different Einstein-Rosen charts. Natural boundary conditions in a scattering problem are --- like gauge theory --- solutions that look flat in the ingoing or outgoing regions:
\begin{equation}
    E^{\mrin}_{i \, a}(x^-< x_i^-) = \delta_{ia}, \qquad E^{\mrout}_{i \, a}(x^- > x_f^-) = \delta_{ia}.
\end{equation}
In general, the solutions will then be linear in the other region:
\begin{equation}
    E^{\mrin}_{i \, a} (x^- > x_f^-) = b^{\mrin}_{i \, a} + c_{i \, a} x^-, \qquad E^{\mrout}_{i \, a} (x^-< x_i^-) = b^{\mrout}_{i\, a} + c_{i \, a} x^-.  \label{sec1:Easymp}
\end{equation}
The matrices $b_{i\, a}, c_{i\, a}$ encode the displacement and velocity memory effect of matter passing through the plane wave. The matrix $c$ does not have an in/out label because the conservation of the Wronskian of the solutions to the  differential equation \eqref{sec1:VielEq} ensures that $c^{\mrin} = c^{\mrout}$. The deformation tensor $\sigma_{ab}$ also encodes memory in the following way 
\begin{gather}
    \sigma_{ab}^{\mrin}(x^- < x_i^-) = 0, \qquad \sigma_{ab}^{\mrin}(x^- > x_f^-) = - (b^{\mrin} + c x^-)^{-1} c, \\ 
    \sigma_{ab}^{\mrout}(x^- > x_f^-) = 0, \qquad \sigma_{ab}^{\mrout}(x^- < x_i^-) = -(b^{\mrout} + c x^-)^{-1} c.
\end{gather}
Overall, the scaling is as $1/x^-$ as $x^- \rightarrow  \pm \infty$ assuming $c \neq 0$. 

We will often refer back to the $x^-$ large-distance scaling~\cite{Cristofoli:2025esy} of various geometric quantities on plane wave background, which are collected in Table \ref{sec1:scaleTable}. As for gauge theory, we will assume that the memory matrix $c$ is non-zero and considered `strong' --- non-negligible compared to other scales in the problem. `Miraculous' plane waves~\cite{Zhang:2024uyp} where $c=0$ will be considered in Section \ref{sec:pertGR}.
\begin{table}
    \centering
    \begin{tabular}{c || c | c | c}
        & $x^- \rightarrow -\infty$ & $x^- \rightarrow + \infty$ &  \\ \hline
        $E^{\mrin}_{i \, a}$ & $ 1$ & $ c x^-$ &\\
        $E^{\mrin \, i}_a$ & $ 1$ & $c^{-1} (x^-)^{-1}$&\\
        $\sigma^{\mrin}_{ab}$ & 0 &$  (x^-)^{-1}$& \eqref{sec1:deform} \\
        $\gamma^{\mrin}_{ij}$ & $1$ & $c^2 (x^-)^2$ & \eqref{sec1:ERmet} \\
        $\gamma^{\mrin \, ij}$ & $ 1$ & $c^{-2} (x^-)^{-2}$ & \\
        $F^{\mrin\, ij}$ & $ x^-$ & $ c^{-2}$ & \eqref{sec1:Fdef}\\
        $\det{E^{\mrin}}$ & 1 & $\det(c) (x^-)^2$&
    \end{tabular}
    \caption{The asymptotic-in-$x^-$ behaviour of useful geometric quantities in a generic gravitational plane wave, ignoring the tensor structure of the quantities. The behaviour for the outgoing gauge are the same, with $+ \infty$ and $-\infty$ swapped.} \label{sec1:scaleTable}
\end{table}

\paragraph{Scalars.} We will consider scalar fields of mass $m$ traversing the background. The solution to the scalar equation of motion on a gravitational plane wave in Brinkmann coordinates
\begin{equation}
    \nabla_{\mu} \nabla^{\mu} \Phi(x) = \Big(2 \partial_+ \partial_- + H_{ab}(x^-) x^a x^b \partial_+^2 - \partial_a \partial^a\Big) \Phi(x) = m^2 \Phi(x)
\end{equation}
is given by~\cite{Friedlander:2010eqa,Ward:1987ws}
\begin{equation}
    \Phi(x) = \frac{1}{\sqrt{\det (E(x^-))}} e^{\im \phi_k} \label{sec1:scalSol}
\end{equation}
where the phase $\phi_k$ (which will be distinguished from the gauge theory variant by context) is
\begin{equation}
    \phi_k \coloneqq k_+ x^+ + k_i E^i_{\, a}(x^-) x^a + \frac{k_+}{2} \sigma_{ab}(x^-) x^a x^b + \frac{k_i k_j}{2k_+} F^{ij}(x^-) + \frac{m^2}{2k_+} x^-.
\end{equation}
The object $F^{ij}(x^-)$ is the integral of the inverse of the Einstein-Rosen transverse metric
\begin{equation}
    F^{ij}(x^-) = \int^{x^-} \gamma^{ij}(s) \d s. \label{sec1:Fdef}
\end{equation}
Again, $(k_+, k_{\perp})$ are the free components of the on-shell momentum. As in gauge theory, we can choose for this solution to satisfy natural initial conditions by choosing a certain Einstein-Rosen chart for the vielbeins.  The initial conditions of interest are again
\begin{equation}
    \Phi^{\mrin}_k (x^- < x_i^-) = e^{\im k \cdot x}, \qquad \Phi^{\mrout}_k(x^- > x_f^-) = e^{\im k \cdot x }.
\end{equation}
These are solved by adding appropriate superscripts to the vielbeins and related quantities, with solutions 
\begin{equation}
    \Phi^{\mrin}_k (x) =  \frac{1}{\sqrt{\det (E^{\mrin}(x^-))}} e^{\im \phi^{\mrin}_k}, \qquad \Phi^{\mrout}_k (x) =  \frac{1}{\sqrt{\det (E^{\mrout}(x^-))}} e^{\im \phi^{\mrout}_k}. \label{sec1:gInOut}
\end{equation}
Note that due to the dependence on the transverse vielbeins, the solution \eqref{sec1:scalSol} will in general have some singularities. The prefactor $\det (E)^{1/2}$ means that an incident wave profile $e^{i k \cdot x}$ will not look like $e^{i k' \cdot x}$ after passing through the wave. In particular, there will be an overall $(x^-)^{-1}$ fall-off as seen after tracking the large-$x^-$ behaviour using Table \ref{sec1:scaleTable}. Therefore, a wave with initial momentum $k$  will not have a definite momentum after the wave (in sharp contrast to the gauge theory case, cf. \eqref{sec1:dMom}), though it can still be decomposed into outgoing Fourier modes~\cite{Garriga:1990dp,Adamo:2017nia,Cristofoli:2025esy}. In fact, the gravitationally dressed momentum $K_{\mu}(x) = \d \phi_k$ is given by 
\begin{multline}
    K_{\mu}\, \d x^{\mu} = k_+ \d x^+ + \big(k_i E^i_a + k_+ \sigma_{ab} x^b \big) \d x^a+ \Bigg( \frac{k_+}{2} \dot{\sigma}_{bc} + k_i \dot{E}^i_b x^b + \frac{k_i k_j}{2k_+} \gamma^{ij} \Bigg) \d x^-  \\ 
    + \frac{m^2 }{2k_+} \d x^-. \label{sec1:gDMom}
\end{multline}
The dependence on $x^a$ in this expression means that this has no definite value asymptotically. This asymptotic behaviour of $\Phi(x)$ was analysed carefully in~\cite{Cristofoli:2025esy}.

\paragraph{Photons.} The linearised equation of motion for the gauge connection on a gravitational plane wave in Lorenz and lightfront gauge is 
\begin{equation}
    g^{\rho \sigma} \nabla_{\rho} \nabla_{\sigma} a_{\mu} = 0, \qquad \partial_{\mu} a^{\mu} = 0 = a_{\mu} n^{\mu}.
\end{equation}
These can again be constructed from the massless scalar solution $\Phi = (\det E)^{1/2} e^{\im \phi_k}$ via a spin-raising operator~\cite{MasonSpin,Adamo:2017nia,Araneda:2022lgu}, where we recall the massless version of the phase:
\begin{equation}
    \phi_k = k_+ x^+ + k_i E^i_{\, a}(x^-) x^a + \frac{k_+}{2} \sigma_{ab}(x^-) x^a x^b + \frac{k_i k_j}{2k_+} F^{ij}(x^-) .
\end{equation}
The spin-1 solution is then
\begin{equation}
    a_{\mu} = \mathcal{E}_{\mu}(x) \, \Phi(x), \qquad \mathcal{E}_{\mu} \coloneqq \epsilon_a \delta^{a}_{\mu} + \Bigg( \frac{k_j}{k_+} E^j_a + \sigma_{ab} x^b \Bigg) \epsilon^a \, n_{\mu}.
\end{equation}
The vector $\mathcal{E}_{\mu}$ is known as the dressed polarisation vector and obeys $\mathcal{E}_{\mu} K^{\mu} = 0$. Again, by choosing the appropriate boundary conditions for the vielbeins we can write down the ingoing and outgoing solutions
\begin{equation}
    a^{\mrin}_\mu = \mathcal{E}^{\mrin}_{\mu} (x) \Phi^\mrin(x), \qquad a^\mrout_\mu = \mathcal{E}^{\mrout}_{\mu} (x) \Phi^{\mrout}(x). 
\end{equation}

\paragraph{Gravitons.} The linearised Einstein equation for a metric fluction $h_{\mu \nu}$ in transverse-traceless gauge on a gravitational plane wave is 
\begin{equation}
    \nabla_{\sigma} \nabla^{\sigma} h_{\mu \nu} - 2 R^\rho_{\mu \nu \sigma} h^{\sigma}_\rho = 0.
\end{equation}
Further imposing light-cone gauge $h_{\mu \nu} n^{\mu} =0$, the spin-2 solution can be written in terms of the massless scalar and spin-1 solutions using the spin-raising operator again
\begin{equation}
    h_{\mu \nu} (x) \coloneqq \mathcal{E}_{\mu \nu}(x) \, \Phi(x)= \Bigg( \mathcal{E}_{\mu} (x)\mathcal{E}_{\nu} (x) - \frac{i}{k_+} \epsilon_a \epsilon_b\, \sigma^{ab}(x) \,n_\mu n_\nu \Bigg)\, \Phi(x). \label{sec1:s2sol}
\end{equation}
The transverse polarisation $\epsilon_a$ is chosen to be null with respect to $\delta_{ab}$. The term proportional to $1/k_+$ is commonly known as the `tail' term. This is due to its role in the Green's function on a plane wave background decomposed in terms of graviton modes~\cite{Adamo:2022qci,Cristofoli:2025esy}. There, this term generates the tail effect --- support of the retarded Green's function in the interior of its past null cone~\cite{Harte:2013dba}. 

As before, ingoing and outgoing boundary conditions are specified with the appropriate superscripts --- $h^{\mrin}_{\mu \nu}, h^{\mrout}_{\mu \nu}$ --- each behaving like a graviton on flat space in the ingoing and outgoing regions respectively.

\subsubsection{Self-dual gravitational plane waves}

As in gauge theory, it will be interesting for us to interpret our results also for self-dual gravitational backgrounds. These are backgrounds where the Weyl tensor is self-dual. In gravity, a self-dual plane wave is described by a single function determining the profile 
\begin{equation}
    H_{ab}(x^-) = \begin{pmatrix}
        \dot{f}(x^-) & - \im \dot{f}(x^-) \\
        - \im \dot{f}(x^-) & - \dot{f}(x^-)
    \end{pmatrix}, \label{sec1:sdGravProf}
\end{equation}
for a sandwich function $\dot{f}(x^-)$. Equivalently in terms of complex coordinates $(z, \bar{z})$ this is $H_{ab}  x^a x^b = \dot{f}(x^-) \bar{z}^2$. Assuming ingoing boundary conditions $E^{\mrin}_{i \, a}(x^- < x_i^-) = \delta_{ia}$ the matrix of vielbeins and its determinant is is 
\begin{equation}
    E_{i \, a} = \begin{pmatrix}
        1 + \int^{x^-} f(s) \d s & - \im \int^{x^-} f(s) \d s \\
        - \im \int^{x^-} f(s) \d s & 1 - \int^{x^-} f(s) \d s
    \end{pmatrix}, \qquad \det(E_{i \, a})  = 1. \label{SDViels}
\end{equation}
The fact that the determinant is constant in $x^-$ is one of the features that distinguishes self-dual plane waves from the non-chiral counterparts. Importantly, we don't have a problem of the focusing of null geodesics and the associated singularities. Nevertheless, these waves will carry velocity and displacement memory as can be seen in the form of $E_{i \, a}$. The transverse Einstein-Rosen metric is 
\begin{equation}
    \gamma_{ij} = \mathbbm{1}_{ij} + \begin{pmatrix}
        2 \int^{x^-} f(s) \d s & - 2 \im \int^{x^-} f(s) \d s \\
        - 2 \im \int^{x^-} f(s) \d s & 2 \int^{x^-} f(s) \d s
    \end{pmatrix} \label{SDERmet}
\end{equation}
making the whole metric in Einstein-Rosen coordinates 
\begin{equation}
    \d s^2 = 2 \d X^+ \d X^- - \d Z \d \bar{Z} + 2 \d \bar{Z}^2 \int^{x^-} f (s)  \, \d s .
\end{equation}
The deformation tensor for a self-dual plane wave is
\begin{equation}
    \sigma_{ab}(x^-) = \begin{pmatrix}
        - f(x^-) & i f(x^-) \\
         i f(x^-) & f (x^-) \label{SDdeform}
    \end{pmatrix}.
\end{equation}
When this is contracted with a negative helicity (anti-self-dual) transverse polarisation vector $\epsilon^{(-)}_a$, we have $\sigma_{ab}\epsilon^{(-)\, a} = 0$. Hence the tail term in the spin-2 solution \eqref{sec1:s2sol} vanishes. This should be compared to the discussion in~\cite{Adamo:2022mev} where the free-field solutions are also constructed but using twistor methods.

\section{Leading soft theorems on plane wave backgrounds} \label{sec:soft}

In this section we consider tree-level scattering amplitudes at all multiplicities on plane wave backgrounds in electromagnetism, Yang-Mills and gravity, with a soft photon, gluon or graviton. As in flat space, amplitudes with soft massless mediators develop singularities. However, the origin and structure of these singularities will differ depending on the background field. In the derivation of these results we will use the Feynman rules for these theories on the background (which can derived using the perturbiner (or AFS) method) summarised in Appendix \ref{app:Feyn}. 

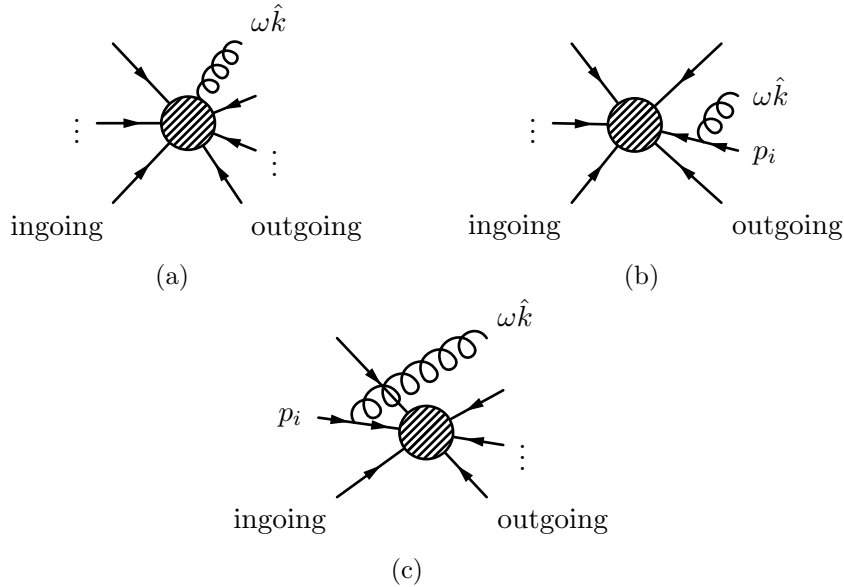
\begin{figure}
\centering
\begin{subfigure}[t]{0.4\textwidth}
    \centering
    \begin{tikzpicture}[baseline=(current bounding box.center)]
        \node {
\bd{ampwithgrav}(60,60)
\fmfset{arrow_len}{2.5mm}
\fmfleft{i1,i2,i3}
\fmfright{o1,o2,o3,o5}
\fmf{fermion}{i1,g}
\fmf{fermion}{i2,g}
\fmf{fermion}{i3,g}
\fmf{fermion}{o1,g}
\fmf{fermion}{o2,g}
\fmf{fermion}{o3,g}
\fmf{gluon}{o5,g}
\fmfv{decor.shape=circle,decor.filled=shaded,decor.size=20}{g}
\fmfv{label=$\text{outgoing}$}{o1}
\fmfv{label=$\vdots$}{o2}
\fmfv{label=$\omega \hat{k}$}{o5}
\fmfv{label=$\vdots$}{i2}
\fmfv{label=$\text{ingoing}$}{i1}
\ed };
\path[use as bounding box] ([shift={(2.5ex,0ex)}]current bounding box.north east) rectangle ([shift={(-2.5ex,-2.5ex)}]current bounding box.south west);
    \end{tikzpicture} 
\caption{}\label{figa}
\end{subfigure}
\begin{subfigure}[t]{0.4\textwidth}
    \centering
    \begin{tikzpicture}[baseline=(current bounding box.center)]
        \node {\bd{ampwithgrav1}(70,60)
        \fmfset{arrow_len}{2.5mm}
        \fmfleft{i1,i2,i3}
        \fmfright{o1,o2,o3,o4}
        \fmf{fermion}{i1,g}
        \fmf{fermion}{i2,g}
        \fmf{fermion}{i3,g}
        \fmf{fermion}{o1,g}
        \fmf{fermion}{o2,g1}
        \fmf{fermion,tension=0.5}{g1,g}
        \fmf{fermion}{o4,g}
        \fmffreeze
        \fmf{gluon}{o3,g1}
        \fmfv{decor.shape=circle,decor.filled=shaded,decor.size=20}{g}
        \fmfv{label=outgoing}{o1}
        \fmfv{label=$p_i$}{o2}
        \fmfv{label=$\omega \hat{k}$}{o3}
        \fmfv{label=$\vdots$}{i2}
        \fmfv{label=ingoing}{i1}
        \ed  };
\path[use as bounding box] ([shift={(2.5ex,0ex)}]current bounding box.north east) rectangle ([shift={(-2.5ex,-2.5ex)}]current bounding box.south west);
    \end{tikzpicture} 
\caption{}\label{figb}
\end{subfigure}
\begin{subfigure}[t]{0.4\textwidth}
    \centering
        \begin{tikzpicture}[baseline=(current bounding box.center)]
            \node {
  \bd{ampwithgrav2}(70,60)
 \fmfset{arrow_len}{2.5mm}
\fmfleft{o1,o2,o4}
\fmfright{i1,i2,i3,o3}
\fmf{fermion}{i1,g}
\fmf{fermion}{i2,g}
\fmf{fermion}{i3,g}
\fmf{fermion}{o1,g}
\fmf{fermion}{o2,g1}
\fmf{fermion,tension=0.5}{g1,g}
\fmf{fermion}{o4,g}
\fmffreeze
\fmf{gluon}{o3,g1}
\fmfv{decor.shape=circle,decor.filled=shaded,decor.size=20}{g}
\fmfv{label=ingoing}{o1}
\fmfv{label=$p_i$}{o2}
\fmfv{label=$\omega \hat{k}$}{o3}
\fmfv{label=$\vdots$}{i2}
\fmfv{label=outgoing}{i1}
\ed };
\path[use as bounding box] ([shift={(2.5ex,2.5ex)}]current bounding box.north east) rectangle ([shift={(-2.5ex,-2.5ex)}]current bounding box.south west);
\end{tikzpicture} 
\caption{}\label{figc}
\end{subfigure}
    \caption{Possible sources to the leading soft singularity of a scattering process. Here the soft particle is treated with outgoing boundary conditions and all momenta point into the diagram. Whether a hard particle is ingoing or outgoing is determined by the sign of $p_{i \, +}$: if $p_{i \, +}>0$, the particle is considered ingoing, and if $p_{i \, +}< 0$ it is considered outgoing.
    In (a), the singularity would come from attaching to internal lines or vertices in the scattering amplitude. In (b) the soft outgoing mediator attaches to an outgoing hard particle. In (c) the soft outgoing mediator attached to an ingoing hard particle. The diagrams (b) and (c) are the ones contributing to the flat space leading soft theorem, but diagrams such as (a) can contribute to soft singularities in the case of soft gravitons.} \label{sec2:softex}
\end{figure}

We will only consider the scattering of scalars in a minimally coupled theory, though the results can be readily generalised to coupling to fields of higher spin. We start by analysing the origins of soft singularities in each of the cases. The usual source of the leading soft singularity on flat space is when the soft particle attaches directly to an external line, with different scenarios depicted in Figure \ref{sec2:softex}. The singularity then comes from the propagator going on shell as the mediator momentum becomes soft, with an infrared divergent factor of $(\epsilon \cdot p_i)^s/k \cdot p_i$ for each external particle with momentum $p_i$ and spin $s$. 

In flat space, this is the only contribution to the leading infrared divergences. The soft particle attaching to an internal line does not contribute as there is no propagator going on shell. The same is the case when the soft particle is part of a higher point vertex. 

On a background, this analysis is a bit less straightforward, as we shall see. The first difference is that the soft particle may gain momentum and cease to be soft as it passes through the background (e.g. \eqref{sec1:massless}). This will then affect which propagators go on shell when it attaches to an external leg. Even worse, the polarisation vector of the particle may itself have a soft singularity  (e.g. \eqref{sec1:s2sol}). This means that just the presence of such a particle attaching anywhere in a scattering amplitude will be infrared divergent. 

\medskip 

In each of the following subsection we consider the soft limit of an $(n+1)$-point amplitude 
\begin{equation}
    \lim_{\omega \rightarrow 0} \omega \, \mathcal{M}_{n+1} (\{p_i\}; \omega \hat{k})
\end{equation}
where $k = \omega \hat{k}$ is the momentum of the soft massless mediator (photon, gluon or graviton), and $\{p_i\}$ are the momenta of the external scalars. When appropriate we will distinguish between the cases of the soft particle having ingoing or outgoing boundary conditions. For the hard scalar particles this is specified by the $+$-component of their momentum. If $p_{i \, +} > 0$ the particle is \emph{ingoing} and therefore it will be represented by a wavepacket with \emph{ingoing} boundary conditions. Conversely, if $p_{i \, +} < 0$, the particle is \emph{outgoing} and will be represented by a wavepacket with \emph{outgoing} boundary conditons. 

\subsection{Electromagnetism}

\begin{figure}
    \centering
    \includegraphics[width=0.55\textwidth]{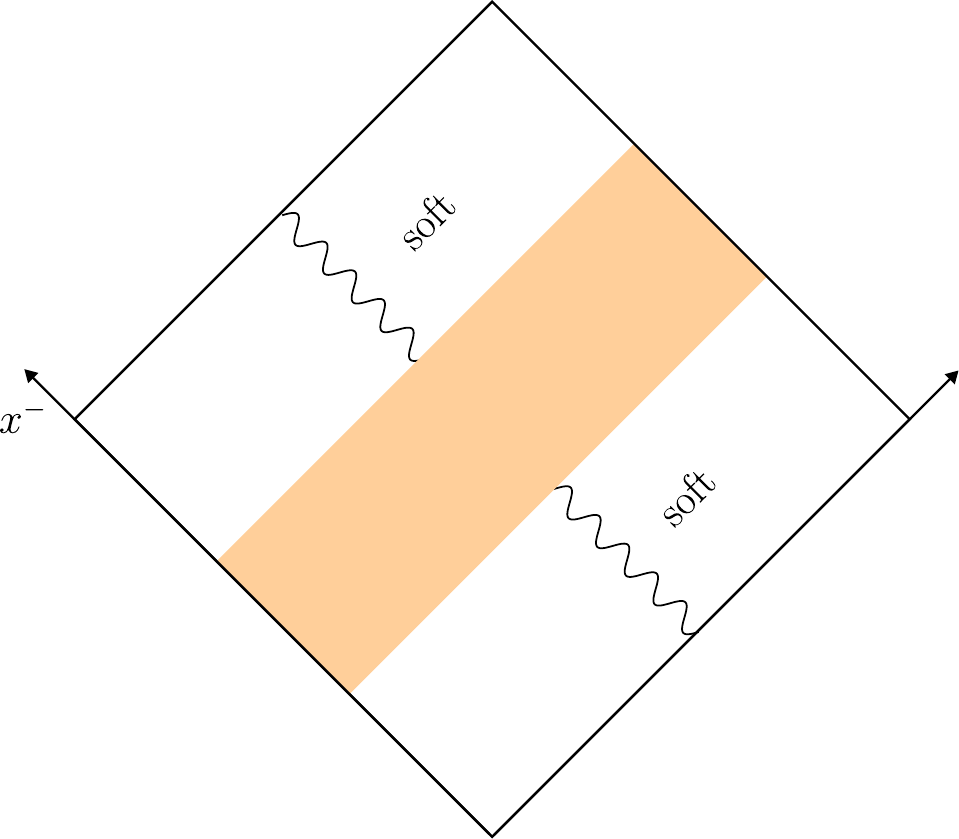}
    \caption{An ingoing soft photon in an electromagnetic plane wave background remains soft in the outgoing region. Therefore soft singularities occur when the ingoing soft photon attaches to both ingoing and outgoing hard particles.}\label{fig:photon}
\end{figure}

In this section we consider the infrared divergences of the scattering of charged, massive scalars with a single soft photon in a gauge theory plane wave background. The photon wavefunction with momentum $k = \omega \hat{k}$ is simply 
\begin{equation}
    a_{\mu}(x) = \epsilon_\mu e^{\im k \cdot x}.
\end{equation} 
The only soft singularity in this instance as $\omega \rightarrow 0$ comes from attaching directly to an external scalar and the propagator going on shell. Soft singularities will therefore only arise from terms in the sum 
\begin{multline}
    \omega \mathcal{M}_{n+1} \supset  e \sum_{p \text{ ingoing}}\omega  Q \int \frac{\d^4 x \, \hat{\d}^4 l}{l^2 - m^2+ \im \epsilon} (P^{\mrin}_\mu(x) + L^{\mrin}_\mu(x)) \epsilon^\mu \, e^{- \im (\phi_p^{\mrin} + k \cdot x - \phi_l^{\mrin})} \mathcal{M}_n(l_p, \ldots) \\
    + e \sum_{p \text{ outgoing}} \omega Q \int \frac{\d^4 x \, \hat{\d}^4 l}{l^2 - m^2  + \im \epsilon} (P^{\mrout}_\mu(x) + L^{\mrout}_\mu(x)) \epsilon^{\mu} \, e^{-\im (\phi_p^{\mrout} + k \cdot x - \phi_l^{\mrout})} \mathcal{M}_n(l_p, \ldots)
\end{multline}
where $\mathcal{M}_n(l_p, \ldots)$ signifies the $n$ point (pure scalar) amplitude with the external leg $p$ taken off-shell with associated momentum $l$. 

\medskip

In this case, the analysis of the ingoing sum and the outgoing sum are very similar as there is no difference in the photon wavefunction. First consider one of the ingoing terms 
\begin{equation}
     \omega\,  e Q \int \frac{\d^4 x \, \hat{\d}^4 l}{l^2- m^2 + \im \epsilon} (P^{\mrin}_\mu (x) - L^{\mrin}_\mu (x) ) \epsilon^\mu \, e^{- \im (\phi_p^{\mrin} + k \cdot x - \phi_l^{\mrin})} \mathcal{M}_n(l_p, \ldots).
\end{equation}
The vector product 
\begin{equation}
    (P^{\mrin}_\mu(x) + L^\mrin_\mu (x)) \epsilon^\mu = \frac{(p_+ + l_+)}{k_+} k_\perp \epsilon^\perp + (p + l )_\perp \epsilon^\perp
\end{equation}
is independent of $l^2, x^\perp$ and $x^+$. The argument of the exponential is proportional to 
\begin{multline}
    \phi_p^{\mrin}(x) + k \cdot x - \phi_l^{\mrin}(x) = (p_+ + k_+ - l_+) x^+ + (p_\perp + k_\perp - l_\perp + (e_k - e_l) \mathsf{A}_{\perp}^{\mrin}) x^{\perp} \\
    + \int^{x^-} \Bigg[ \frac{(p_\perp + e_p A_{\perp}^\mrin(s))^2 + m^2}{2p_+} + \frac{k_\perp^2}{2k_+} - \frac{(l_\perp + e_l A_\perp^\mrin (s))^2 }{2l_+} - \frac{l^2}{2l_+}\Bigg] \d s.
\end{multline}
Through charge conservation (as the background is Cartan-valued) we know that $e_p = e_l \eqqcolon e$. We can also evaluate the integrals in $x^+$ and $x^{\perp}$ straightforwardly to fix the propagator momentum in those directions 
\begin{align}
    l_+ = p_+ + k_+, \\
    l_\perp = p_\perp + k_\perp.
\end{align}
We now start taking the soft limit by rescaling $k_{\mu} = \omega \hat{k}_\mu$, where $\omega$ is a small variable and $\hat{k}_0 = 1$. Expanding to leading order in $\omega$ the leading contribution to the integral is 
\begin{multline}
    2 \omega\,  e Q  \int \d x^- \, P^\mrin_{\mu} (x^-) \epsilon^\mu \\
    \times \exp \Bigg[- \im \int^{x^-} \omega \Big( \frac{\hat{k}^2_\perp}{2 \hat{k}_+} - \frac{2 \hat{k} \cdot (p + e A^\mrin(s))}{2p_+} + \frac{\hat{k}_+}{2p_+^2} (p_{\perp} + e A_{\perp}^\mrin(s))^2  + \frac{m^2 \hat{k}_+ }{2p_+^2}\Big) d s \Bigg] \\
    \times\int \frac{\hat{\d} (l^2)}{2l_+ \, (l^2 - m^2+ \im \epsilon)} \exp \Big[ \frac{ \im (l^2- m^2)}{2l_+} (x^- - y^-)\Big] \mathcal{M}^{y^-}_n(l_p, \ldots).
\end{multline}
Here we have pulled out the $e^{\im l^2 y^-}$ term coming from the next vertex in $\mathcal{M}_n(l_p, \ldots)$ to define the $(y^-)$-dependent subamplitude $\mathcal{M}_n^y(l_p, \ldots)$. Evaluating the last line with the Feynman contour prescription we find 
\begin{equation}
    \int \frac{\hat{\d} (l^2)}{2l_+ \, (l^2 -m^2+ \im \epsilon)} \exp \Bigg[  \frac{\im (l^2- m^2)}{2l_+} (x^- - y^-)\Bigg] = - \frac{ \im}{2l_+} \Theta\Bigg(\frac{y^- - x^-}{2p_+}\Bigg) \exp \Bigg[\frac{\epsilon (x^- - y^-)}{2l_+}\Bigg].
\end{equation}
This means that when $p_+ > 0$ (an ingoing scalar, as we've assumed here), we always have $x^- < y^-$ where $y$ is the spacetime location of the subsequent vertex in the scattering process. Incorporating this into the above expression we have 
\begin{multline}
    - \frac{\im\, \omega\,  e Q  }{p_+} \int_{- \infty}^{y^-} \d x^- \, P^\mrin_\mu \epsilon^\mu \exp \Bigg[\frac{\epsilon (x^- - y^-)}{2p_+} \Bigg] \\
    \times \exp \Bigg[ - \im \int^{x^-} \omega \Big( \frac{\hat{k}^2_\perp}{2 \hat{k}_+} - \frac{2 \hat{k} \cdot (p + e A^\mrin(s))}{2p_+} + \frac{\hat{k}_+}{2p_+^2} (p_{\perp} + e A_{\perp}^\mrin(s))^2  + \frac{m^2 \hat{k}_+ }{2p_+^2}\Big) d s \Bigg]  \\ 
    \times \mathcal{M}^{y^-}_n (p, \ldots), \label{sec2:photfinal}
\end{multline}
where higher order contributions in $\omega$ have now been dropped. In particular, the ingoing momentum into $\mathcal{M}_n^y(l_p, \ldots)$ can be readily changed to  $\mathcal{M}_n^{y^-}(p, \ldots)$ to leading order in $\omega$. Now this integral is of the form 
\begin{equation}
    \omega\,  e Q  \int^{y^-}_{-\infty} \d x^- \, g(x^-)\, \exp \Bigg[ \frac{\epsilon (x^- - y^-)}{2p_+}\Bigg] \times \exp [\im \, \omega f(x^-)],
\end{equation}
where $f(x^- < x_i^-) = A x^-$ and $g(x^- < x_i^-) = B$ for
\begin{equation}
A = - \frac{\hat{k} \cdot p}{p_+}, \qquad B = - \frac{\im}{p_+} \,\epsilon \cdot p \times \mathcal{M}_n^{y^-} (p , \ldots),
\end{equation}
 $\epsilon,p_+ > 0$ and $\epsilon$ is being taken to zero. The values of $A, B$ can be seen by using the fact that $A^\mrin (x^- < x_i^-) = 0$ and the expansion of $\hat{k} \cdot p$ for on-shell momenta written in lightfront coordinates. Using the results of appendix \ref{app:func1} the leading $\omega \rightarrow 0$ behaviour of \eqref{sec2:photfinal} is then
\begin{equation}
    e Q \,\frac{\epsilon \cdot p}{\hat{k} \cdot p}\times  \mathcal{M}_n(p, \ldots).
\end{equation}
Because the prefactor no longer depends on $y$, we here drop the $y$ label from $\mathcal{M}_n(p, \ldots)$.

\medskip
Repeating the same argument for the rest of the ingoing legs and the outgoing legs (which will involve the final integral being over the interval $[y^-, \infty)$) we reproduce the soft theorem for photons on a gauge theory plane wave background.
\begin{tcolorbox}[colback=red!5!white,colframe=red!75!black,title=Soft photon theorem on a gauge theory plane wave~\cite{Ilderton:2012qe,Dinu:2012tj}]\vspace{-0.35cm}
\begin{equation}
    \lim_{\omega \rightarrow 0} \omega \mathcal{M}_{n + 1}(\{p_i\}; \omega \hat{k}) = e \, \Bigg[ \, \sum_{p_i \text{ ingoing}}  Q_i\, \frac{\epsilon \cdot p_i}{\hat{k} \cdot p_i}  + \, \sum_{p_i \text{ outgoing}}  Q_i \,\frac{\epsilon \cdot p_i}{ \hat{k} \cdot p_i} \Bigg] \mathcal{M}_n(\{p_i\})
\end{equation}
\end{tcolorbox}

This is the same as the leading soft photon theorem on flat space. The reason for this is because the photon doesn't couple to the background --- it only sees the hard charged scalars. Additionally, viewing the infrared divergent behaviour as coming from integrals over infinite distances, the short time that charged particles experience a change of momentum during the scattering process is nothing compared to the infinity of time that they've spent with their initial/final momenta.  

\subsection{Yang-Mills} \label{sec:YM}

\begin{figure}[t]
    \centering
    \includegraphics[width=0.55\textwidth]{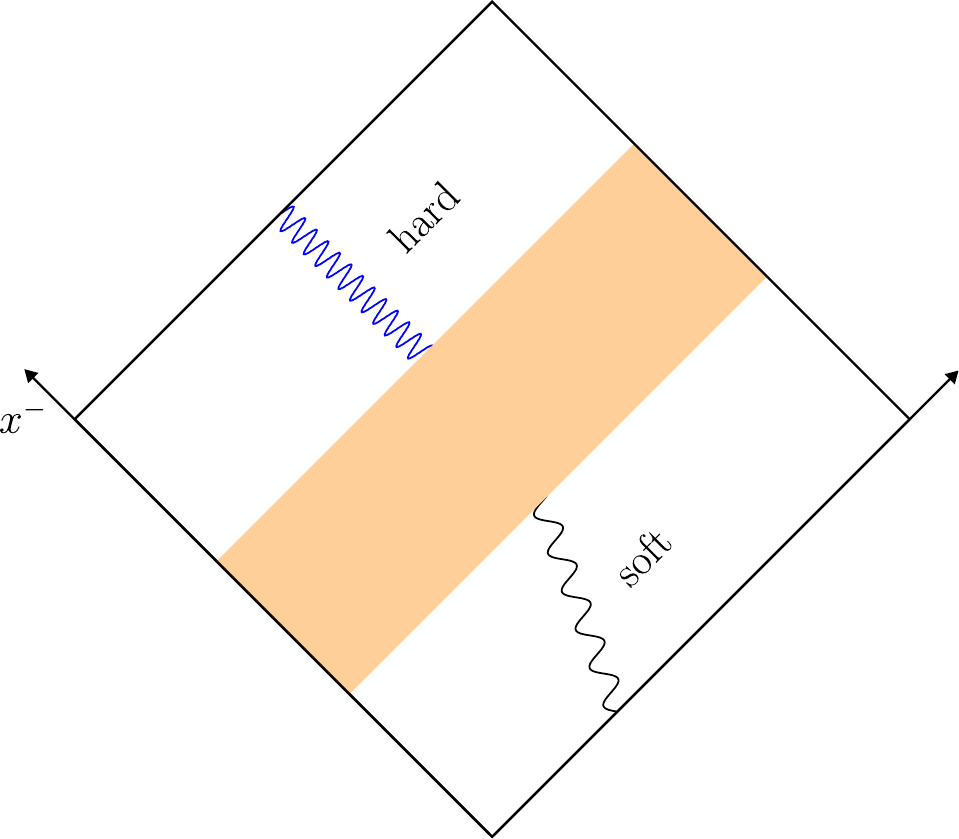}
    \caption{A gluon in a Cartan-valued plane wave background that is soft in the ingoing region becomes infinitely hard in the outgoing region. Intuitively, this is why soft singularities arising from an ingoing soft gluon only happen when it couples to an also ingoing hard particle.} \label{fig:gluon}
\end{figure}

A key distinction between photons and gluons defined on a gauge theory plane wave background is that soft gluons see the background field and, importantly, don't stay soft. See Figure \ref{fig:gluon}. This is perhaps most clear from the dressed gluon momentum 
\begin{equation}
    K_{\mu}(x^-) \, \d x^\mu = k_+ \d x^+ + (k_a + e A_a (x^-)) \, \d x^a +  \frac{1}{2k_+} (k_\perp + e A_\perp(x^-))^2 \, \d x^- 
\end{equation}
where $e$ is the charge of the gluon with respect to the Cartan-valued background generator. For example fixing ingoing boundary conditions on $A_{\perp}$, we see that an ingoing soft gluon with momentum $k_\mu = \omega \hat{k}_{\mu}$ does not stay soft. In fact the leading behaviour in the outgoing region is
\begin{equation}
    K_\mu(x^- > x_f^-) \sim  \frac{e^2 \, |A_{\perp} (x^- > x_f^-)|^2}{2 \omega \hat{k}_+} n_{\mu} + \mathcal{O}(\omega^0).
\end{equation}
For generic memory and a real profile this will always be non-zero\footnote{Notably, it's zero for self-dual plane wave backgrounds. We will return to this in Section \ref{sec:SD}.}. Even further, the dressed polarisation 
\begin{equation}
    \mathcal{E}_\mu (x^-) \, \d x^\mu = \epsilon_a \Bigg( \d x^a + \frac{1}{k_+} (k^a + eA^a) \d x^-\Bigg)
\end{equation}
also has a divergent piece as $\omega \rightarrow 0$. This piece behaves as 
\begin{equation}
    \mathcal{E}_{\mu} (x^- > x^-_f) \sim \frac{e \, \epsilon_a A^a (x^- > x^-_f)}{\omega \hat{k}_+} n_{\mu}. \label{sec2:gludivpol}
\end{equation}
Again, this will generically be non-zero, unless the polarisation is orthogonal to the background profile. One way to view this piece is as this charged, soft gluon being kicked by the plane wave to become entirely collinear with the background. For the rest of the section, we will consider the generic case where both of these leading singularities are non-zero. We will remark on special cases in Section \ref{sec:Comp}.

\medskip 

Having briefly explored the key soft properties of the gluon wavefunction on a plane wave background, we turn to amplitudes. It is still possible to introduce soft gluon states in the asymptotic past or future, and see how singularities appear in the scattering amplitudes. As the polarisation and dressed momenta themselves are now divergent, we can no longer argue that the only singularity will come from an on-shell propagator as in the previous section. We will therefore consider separately the two possible contributions to the soft singularity: spurious terms that occur at every vertex the dressed gluon is attached to, through its dressed polarisation; and from on-shell propagators when the soft gluon attaches directly to an external leg.

\paragraph{Spurious terms are subleading.} Outside of when a propagator becomes on-shell when a soft gluon is attached directly to an external particle, the only other singularities must come directly from the dressed polarisation vector or the dressed momentum. The latter can be ignored by integration by parts at the relevant vertex. The former requires some more careful treatment. Schematically, at each vertex we consider 
\begin{equation}
     \omega \int \d^4 x \, \mathcal{E}_\mu(x) \, e^{\im \phi_{\omega \hat{k}}(x)} \times \mathcal{M}^x(\ldots)
\end{equation}
where $\mathcal{M}^x(\ldots)$ has no further dependence on $\omega$. To this integral, we can now apply a stationary phase approximation as $\phi_{\omega \hat{k}}$ is rapidly oscillating as $\omega \rightarrow 0$ :
\begin{equation}
    \phi_{\omega \hat{k}} (x) = \omega \hat{k}_+ x^+ + (\omega \hat{k}_\perp + e \mathsf{A}_{\perp}) x^\perp + \frac{1}{2 \omega \hat{k}_+} \int^{x^-} \d s \Big[ \omega \hat{k}_\perp + e \mathsf{A}_\perp (s)\Big]^2.
\end{equation}
Applying stationary phase, the leading parts of this integral come from where $\mathsf{A}_\perp (x^-)^2 = 0$. For a real-valued plane wave, this will correspond to $\mathsf{A}_\perp (x^-) = 0$. But these are precisely the points where the dressed polarisation \eqref{sec2:gludivpol} is no longer divergent. The next order is suppressed in $\omega$, and so we find that 
\begin{equation}
    \omega \int \d^4 x \,\mathcal{E}_\mu(x) \, e^{\im \phi_{\omega \hat{k}}(x)} \times \mathcal{M}^x(\ldots) \sim o(\omega^{1/2})
\end{equation}
when taking account of the $\omega^{-1}$ from the dressed polarisation and the $\omega^{1/2}$ suppression coming from the rapidly oscillating phase. We have thus shown that these terms are suppressed relative to the other infrared singularities we study next.

\paragraph{On-shell propagators.} We now restrict to the only other source of a soft singularity: when the gluon attaches directly to an external leg. We will first consider an ingoing gluon (which is soft in the ingoing region, and hard in the outgoing region). We will consider a similar sum to previously 
\begin{multline}
    \omega \mathcal{M}_{n+1}^\mathsf{a} \supset \sum_{p \text{ ingoing}} \omega \, g_{\text{YM}} \int \frac{\d^4 x \,\hat{\d}^4 l}{l^2- m^2 + \im \epsilon} (P^{\mrin}_\mu(x) + L^{\mrin}_\mu(x)) \mathcal{E}^{\mrin \, \mu} (x)\\ \times e^{- \im (\phi_p^{\mrin} + \phi_k^\mrin- \phi_l^{\mrin})} T^{\mathsf{a}}_p \mathcal{M}_n(l_p, \ldots) \\
    + \sum_{p \text{ outgoing}}\omega \, g_{\text{YM}} \int \frac{\d^4 x \, \hat{\d}^4 l}{l^2 - m^2 + \im \epsilon} (P^{\mrout}_\mu(x) + L^{\mrout}_\mu(x)) \mathcal{E}^{\mrin \, \mu} (x) \\ \times e^{-\im (\phi_p^{\mrout} + \phi_k^\mrin - \phi_l^{\mrout})} (T^{\mathsf{a}}_p)^{\ast}\mathcal{M}_n(l_p, \ldots),
\end{multline}
where the superscript on $\mathcal{M}_{n+1}^\mathsf{a}$ denotes that the soft gluon has generator $T^{\mathsf{a}}$. On the right hand side, $T^{\mathsf{a}}_p$ (or $(T^\mathsf{a}_p)^{\ast}$) acts on the index associated with $p$ in $\mathcal{M}_n(l_p, \ldots)$. We will first consider the contribution  from the second line with opposing boundary conditions --- where the ingoing gluon attached to an outgoing scalar ($p_+ < 0$). Ignoring the colour structure for now, consider a single term in this sum 
\begin{equation}
    \omega \, g_{\text{YM}}\int \frac{\d^4 x \, \hat{\d}^4 l}{l^2 - m^2 + \im \epsilon}  (P^{\mrout}_\mu(x) + L^{\mrout}_\mu(x)) \mathcal{E}^{\mrin \, \mu} (x) \, e^{-\im (\phi_p^{\mrout} + \phi_k^\mrin - \phi_l^{\mrout})} \mathcal{M}_n(l_p, \ldots).
\end{equation}
The inner product $(P^{\mrout}_\mu(x) + L^{\mrout}_\mu(x)) \mathcal{E}^{\mrin \, \mu} (x)$ will only depend on $x^-$. The terms in the exponential are now proportional to 
\begin{multline}
    \phi_p^\mrout + \phi_k^\mrin - \phi_l^\mrout = (p_+ + k_+ - l_+) x^+ + (p_\perp + k_\perp  + e_k a_{\infty \, \perp}- l_\perp + (e_p + e_k - e_l)\mathsf{A}^{\mrout})x^\perp \\
    + \int^{x^-} \d s \Bigg[ \frac{(p_\perp + e_p \mathsf{A}_\perp^\mrout (s))^2}{2p_+} + \frac{(k_\perp + e_k \mathsf{A}_\perp^\mrin (s))^2}{2k_+} - \frac{(l_\perp + e_l \mathsf{A}_\perp^\mrout (s))^2}{2l_+}\Bigg] \\
    + \Bigg(\frac{m^2}{2p_+} - \frac{ l^2}{2l_+} \Bigg)x^-.
\end{multline}
Here we have used the fact that $\mathsf{A}^{\mrin} = \mathsf{A}^{\mrout} + a_{\infty}$ as in \eqref{sec1:gaugeMem}. 
This means that we can evaluate all our $x^{\perp}, x^+$ integrals, setting 
\begin{align}
    l_+ &= p_+ + k_+, \label{GaugeL1} \\
    l_\perp &= p_\perp + k_\perp + e_k a_{\infty \, \perp} \label{GaugeL2}
\end{align}
making use of charge conservation $e_l = e_p + e_k$. Substituting these in, neglecting $k_+$ terms relative to $p_+$ in the soft limit, and evaluating the integral for the propagator  takes us to the leading contribution coming from 
\begin{equation}
    - \frac{\im \,\omega \, g_{\text{YM}}}{2p_+} \int_{y^-}^{\infty} \d x^- \,  (P_\mu^\mrout(x) + L_{\mu}^\mrout(x))\mathcal{E}^{\mrin \, \mu}(x) \, e^{- \im \mathcal{V}(k, p) - \epsilon (x^- - y^-)/2p_+} \, \mathcal{M}_n^y(l_p, \ldots). \label{sec2:gluOutFin}
\end{equation}
The dressed momenta are evaluated on the localised values of $l_{+, \perp}$ in (\ref{GaugeL1}, \ref{GaugeL2}). Note that the integration interval is $[y^-, \infty)$ because we assume for outgoing momenta that $p_+ < 0$ which changes the sign of the arguement of the Heaviside $\Theta$-function in \eqref{app:propInt}. As before, we signify that $\mathcal{M}_n^y(l_p, \ldots)$ contains an integral over $y$ with the superscript. Here the Volkov exponent is defined as 
\begin{multline}
    \mathcal{V}(k, p) \coloneqq \int^{x^-} \d s \Bigg[ \frac{(p_\perp + e_p A_\perp^\mrout (s))^2}{2p_+} + \frac{(k_\perp + e_k A_\perp^\mrin (s))^2}{2k_+}\\  - \frac{(p_\perp + k_\perp + e_k a_{\infty\, \perp}+ (e_p + e_k ) \mathsf{A}_\perp^\mrout (s) )^2}{2(p_+ + k_+)} \Bigg] \\+ \Bigg( \frac{m^2}{2p_+} - \frac{m^2}{2(p_+ + k_+)}\Bigg)x^-.
\end{multline}
Even without expanding entirely in small $\omega$, we can see that this exponent has a rapidly oscillating phase coming from the term 
\begin{equation}
    \int^{x^-} \d s \frac{e^2_k |\mathsf{A}_\perp^\mrin (s)|^2}{2 \omega \hat{k}_+}
\end{equation}
in $\mathcal{V}(k, p)$.
Via stationary phase, the leading contribution to the integral are therefore the sets of points where this is stationary, i.e. 
\begin{equation}
    |\mathsf{A}_\perp^\mrin (x^-)|^2 = 0. \label{zeroset}
\end{equation}
All contributions not in this set will be supressed in $\omega$, contributing at least at $\mathcal{O}(\omega^{1/2})$. 
The regions in $x^-$ where \eqref{zeroset} is true for a generic sandwich profile fall into two classes: isolated points or intervals. For the isolated points, we can apply the same argument as before. The integral \eqref{sec2:gluOutFin} is entirely localised to those solutions with an additional factor of $\omega^{1/2}$ from the stationary phase. These contributions are therefore $\mathcal{O}(\omega^{3/2})$ and don't contribute at leading order. We therefore only need to consider intervals. 

With the assumption that we have memory, so $|\mathsf{A}_\perp^\mrin(x^- > x_f^-)| \neq 0$, these intervals can be collected into 
\begin{equation}
    I =  \bigcup_{i = 1}^m I_i, \qquad I_1 = (-\infty, x^-_i], \quad I_j \subset [x_i^-, x^-_f] \quad  \forall j\neq 1. \label{sec2:Iint}
\end{equation}
On these intervals, the integrals we're considering collapse to the ones in electromagnetism as $A^\mrin_\perp (x^-) = 0$ and the particles have no background to be dressed by. The leading contribution in $\omega$ is therefore 
\begin{multline}
    - \frac{\im \, \omega \, g_{\text{YM}}}{p_+}  \int_{I\cap [y^- ,\infty) }\d x^- \,  (p - e_p a_\infty)\cdot \epsilon \, \exp\Bigg[\frac{\epsilon(x^- - y^-)}{2p_+}\Bigg] \exp \Bigg[ - \frac{\im \omega \hat{k}\cdot (p - e_p a_\infty) x^-}{p_+} \Bigg] \\ \times \mathcal{M}_n^y(p , \ldots). \label{sec2:gluOutFinFin}
\end{multline}
A few notes on getting to this line. We recall the definition of $a_\infty = \mathsf{A}^\mrin(x^-) - \mathsf{A}^\mrout (x^-)$ from \eqref{sec1:gaugeMem}. This is a constant that encodes the memory of the background. Since we are only considering the intervals where $A^\mrin(x^-) = 0$, we must have $A^\mrout(x^-) = - a_\infty$. This means that the dressed momentum $P^\mrout \rightarrow p + e_p a_\infty$, and the same goes for the $l_p$ momentum feeding into the subamplitude $\mathcal{M}_n^y$ defined in (\ref{GaugeL1}, \ref{GaugeL2}). We've also used this to rewrite the Volkov exponent in the second exponential. Lastly, we are integrating over $I \cap [y^-, \infty)$ imposed on us by the conditions for the Feynman propagator. This fact is key because $I \cap [y^-, \infty)$ is \emph{compact} coming from \eqref{sec2:Iint}. This means that the $x^-$ integral is over a bounded set, and so we can exchange the $\omega \rightarrow 0$ limit and the integration, which gives zero due to the overall factor of $\omega$. In essence
\begin{equation}
\lim_{\omega \rightarrow 0} \omega \, g_{\text{YM}} \int \frac{\d^4 x \, \hat{\d}^4 l}{l^2 + \im \epsilon}  (P^{\mrout}_\mu(x) + L^{\mrout}_\mu(x)) \mathcal{E}^{\mrin \, \mu} (x) \, e^{-\im (\phi_p^{\mrout} + \phi_k^\mrin - \phi_l^{\mrout})} \mathcal{M}_n(l_p, \ldots) = 0
\end{equation}
Therefore, in contrast to the soft photon, attaching an \emph{ingoing} soft gluon to an \emph{outgoing} scalar does not contribute to the leading soft singularity. Intuitively, this is because the ingoing soft gluon is just not soft in the outgoing region and so there is no propagator to go on shell where this particle is attached.

\medskip 

Let us now consider attaching the ingoing gluon to an ingoing $p_+ > 0$ leg, which has contribution: 
\begin{equation}
    \omega \, g_{\text{YM}} \int \frac{\d^4 x \, \hat{\d}^4 l}{l^2 + \im \epsilon} (P^{\mrout}_\mu(x) + L^{\mrout}_\mu(x)) \mathcal{E}^{\mrin \, \mu} (x) \, e^{-\im (\phi_p^{\mrout} + \phi_k^\mrin - \phi_l^{\mrout})} \mathcal{M}_n(l_p, \ldots).
\end{equation}
The previous argument follows through, up to \eqref{sec2:gluOutFinFin} which is now 
\begin{equation}
    - \frac{\im \, \omega \, g_{\text{YM}}}{p_+} \int_{I \cup (-\infty, y^-]} \d x^- \, (p \cdot \epsilon) \, \exp \Bigg[ \frac{\epsilon (x^- - y^-)}{2p_+} \Bigg]\, \exp \Bigg[ - \frac{ i \omega (\hat{k} \cdot p ) x^-}{2p_+}\Bigg] \times \mathcal{M}_n^y(p, \ldots).
\end{equation}
Here $I$ is the same union of intervals as before. This time since $p_+>0$, the Feynman prescription means that the integral is initially over $[- \infty, y^-]$. Additionally, since the dressed momentum is $P^\mrin(x)$ in this case, the dressing disappears on the intervals $I$. The resulting integral is infinite, and we need to be more careful in swapping limits. Adapting the result in Appendix \ref{app:func1} to this integral, we have the leading $\omega \rightarrow 0$ behaviour 
\begin{equation}
    \frac{p \cdot \epsilon}{\hat{k} \cdot p} \times \mathcal{M}_n (p , \ldots).
\end{equation}
Note that if we had assumed zero memory ($A^\mrin(x) = A^\mrout(x)$) in any of the above analysis, there would have been no distinction between attaching to  ingoing or outgoing scalars and they would have contributed equally. This is of course because that is the only real case where soft gluons remain soft and can therefore contribute to singularities in both asymptotic regions. The key observation here is that this is not generally the case. In Section \ref{sec:pert} we will explore the case of zero memroy. 

\medskip 

A corresponding analysis can be carried out when we have an \emph{outgoing} soft gluon. In that case, all of the terms where it attaches to an external \emph{ingoing} scalar are subleading. In contrast, attaching to an \emph{outgoing} scalar gives the usual soft factor. 

\medskip 

The leading soft gluon theorem on a Cartan-valued plane wave with memory therefore has two variants. 
\begin{tcolorbox}[colback=red!5!white,colframe=red!75!black,title=Soft gluon theorem on a gauge theory plane wave background]\vspace{-0.35cm}
    \begin{align}
    \lim_{\omega \rightarrow 0} \omega \mathcal{M}_{n+1}^\mathsf{a} (\{ p_i\}; \omega \hat{k}^\mrin) &= g_{\text{YM}}\sum_{p_i \text{ ingoing}} \frac{\epsilon \cdot p_i}{\hat{k} \cdot p_i} \, T^{\mathsf{a}}_{p_i} \, \mathcal{M}_n(\{p_i\}), \label{gluon1} \\
    \lim_{\omega \rightarrow 0} \omega \mathcal{M}_{n+1}^\mathsf{a} (\{ p\}; \omega \hat{k}^\mrout) &= g_{\text{YM}} \sum_{p_i \text{ outgoing}} \frac{\epsilon \cdot p_i}{\hat{k} \cdot p_i}\,(T^{\mathsf{a}}_{p_i})^{\ast} \,\mathcal{M}_n(\{p\})\label{gluon2}
\end{align}
\end{tcolorbox}

Here the distinction between ingoing and outgoing gluons is crucial in establishing the result.

\subsection{Gravity}

We now turn to gravitational plane wave backgrounds. Here we will consider both soft photons and soft gravitons. Gravitational backgrounds are distinct from the above two examples. In contrast to photons which don't interact with the background, both photons and gravitons are dressed by the background field. And in contrast to soft gluons which become hard upon passing through the wave, soft photons and gravitons stay soft. The latter can be seen by looking at the dressed momentum \eqref{sec1:gDMom} in a gravitational plane wave background 
\begin{multline}
    K_{\mu}(x) \, \d x^\mu = k_+ \d x^+ + \Bigg(\frac{k_+}{2} \dot{\sigma}_{bc} x^b x^v + k_i \dot{E}^i_{\, a} x^b + \frac{k_i k_j}{2k_+} \gamma^{ij} \Bigg) \d x^- \\ 
    + (k_i E^{i}_a + k_+ \sigma_{ab} x^b) \d x^a
\end{multline}
where all terms are linear in the momentum scale of the graviton.
One may worry that the spatial dependence of the dressed momentum will make an initially soft momentum hard in some asymptotic regions. But taking a careful $r \rightarrow \infty$ limit, keeping the position $u = t - r$ along $\mathcal{I}^+$ fixed, the dressed momentum for an ingoing particle has asymptotic behaviour 
\begin{equation}
    K^\mrin_\mu (u, r, \hat{x}) \, \d x^{\mu} = k_+ \d x^+ - \frac{k_+ (1 + \hat{x}^3)}{1- \hat{x}^3} \d x^- + \frac{\sqrt{2}k_+ \hat{x}_b}{1 - \hat{x^3}} \d x^b,
\end{equation}
where $x_a =r \hat{x}_a$ for $a = 1, 2, 3$. It may look like the dressed momentum only depends on $k_+$, but the whole wavefunction $\Phi(x)$ still depends on the transverse components of the momentum. However unlike a plane wave $e^{\im k \cdot x}$ wavefunction, an ingoing scalar solution will not localise onto one point on the celestial sphere as described in~\cite{Cristofoli:2025esy}. So, whilst the momentum does remain soft, the effect of the gravitational plane wave on initial momentum eigenstates is to spread them out across a range of momenta. 

Another way to see the same thing is using a Bogoliubov transformation, relating ingoing creation and annihilation operators to outgoing ones. Here 
\begin{gather}
    \Phi^\mrin_k(x) = \int \d^3 l_{\text{on-shell}} \, \alpha_{k, l} \,\Phi^\mrout_l (x), \\ 
    a^\mrout(k) = \int \d^3 l_{\text{on-shell}}  \,\alpha_{k, l}\, a^\mrin(l),  \label{sec2:BogoCre}
\end{gather}
where the on-shell measure is over momenta satisfying $l^2 = 0$ for massless momenta, encoded by $\d^3 l_{\text{on-shell}} = \d^4 l \,\delta(l^2)$. The Bogoliubov transformation parameters can be evaluaed via the Klein-Gordon inner product on a constant lightfront slice $\Sigma$ at $x^- = x_0^-$, as in~\cite{Garriga:1990dp,Adamo:2017nia}. This slice is best chosen before we hit a coordinate singularity in the ingoing and outgoing vielbeins corresponding to the focussing of null geodesics. Here we choose $x_0^- < x_i^-$. Then 
\begin{align}
    \alpha_{k, l} &= \int_{\Sigma} \d^3 x \Phi^\mrin_k (x) \bar{\Phi}^\mrout_l (x) \\ 
    &= \frac{4 \pi}{\im} \delta(k_+ - l_+) \frac{e^{- \im(s_{k, l} + r_{k, l})}}{\sqrt{|E^\mrout(x_0)|}}
\end{align}
where the memory-dependent quantities $s_{k, l}$ and $r_{k, l}$ are 
\begin{gather}
    s_{k, l} = \frac{l_i l_j}{2l_+} F^{ij}_{\mrout}(x_0) - \frac{k_i k_j}{2k_+} x_0^-, \\
    r_{k, l} = - \frac{1}{2l_+} (k_a - l_i E^{\mrout \, i}_a (x_0)) (\sigma^\mrout(x_0)^{-1})^{ab} (k_b - l_j E^{\mrout \, j}_b (x_0)).
\end{gather}
Imposing that $k$ is soft in these expression imposes via the saddle-point approximation that the leading contribution to \eqref{sec2:BogoCre} is also when $l$ is a soft momentum. However, once $l$ is assumed to be soft, the transverse momenta still need to be integrated over. This is the statement that the outgoing momentum is `smeared' in the transverse direction. 

\subsubsection{Soft photon on gravitational plane waves} \label{sec:photonGR}

\begin{figure}
    \centering
    \includegraphics[width=0.55\textwidth]{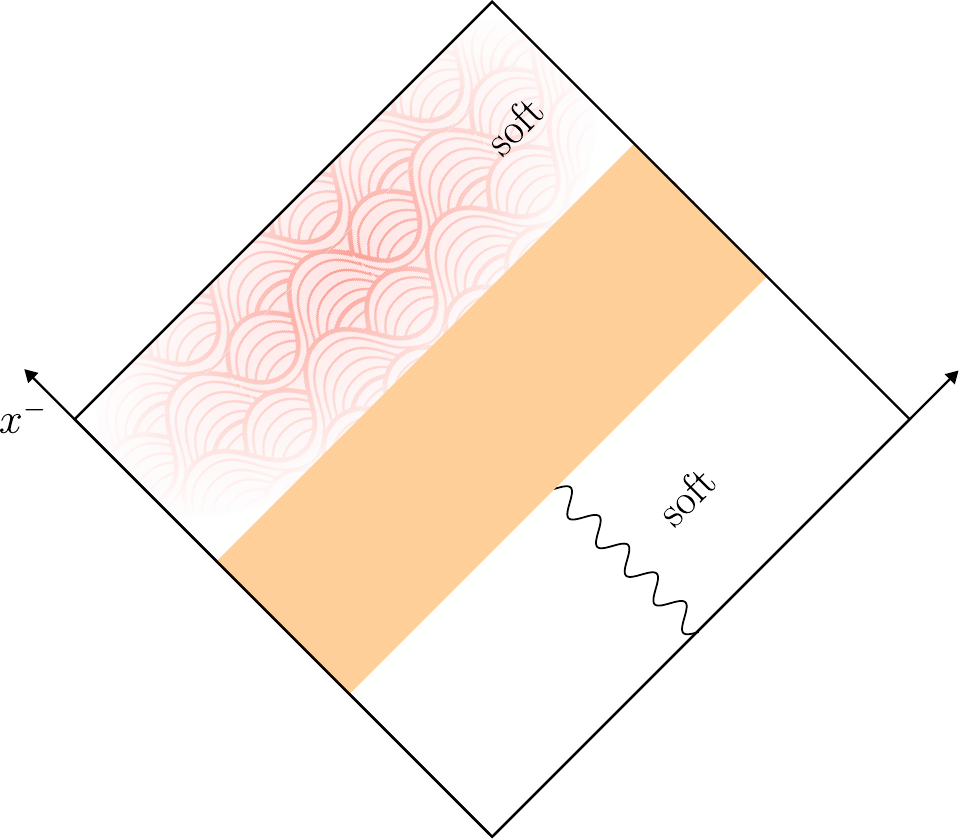}
    \caption{A photon on a gravitational plane wave background that is soft in the ingoing region does not look like a soft Fourier mode in the outgoing region. Instead, it `spreads' out and the analysis of soft singularities when it couples to outgoing hard particles depends on the fall-off conditions.} \label{fig:photonGR}
\end{figure}

We will first consider the soft singularity arising from a soft photon on a gravitational plane wave interacting with charged scalars. Since the dressed polarisation vector has no inherent soft singularities, we can apply the same arguments as in the previous subsections: the only soft singularities will arise when a soft photon attaches to an external leg and a propagator goes on-shell. First considering an ingoing photon, the leading terms will come from 
\begin{multline}
    \omega \mathcal{M}_{n+1} \supset \sum_{p \text{ ingoing}} \omega\, e Q  \int \frac{\d^4 x \, \hat{\d}^4 l\, |E^\mrin (x)|^{- 3/2} }{ (l^2 - m^2+ \im \epsilon)} (P^\mrin_\mu(x) + L^\mrin_\mu(x)) \mathcal{E}^{\mrin \, \mu}(x) \\ \times e^{- \im (\phi_p^\mrin + \phi_k^\mrin - \phi_l^\mrin)} \mathcal{M}_n(l_p, \ldots) \\
    + \sum_{p \text{ outgoing}}\omega\, e Q \int \frac{\d^4 x \, \hat{\d}^4 l\, |E^\mrout(x)|^{-1} }{ |E^\mrin(x)|^{1/2}(l^2 - m^2 + \im \epsilon)} (P_\mu^\mrout (x) + L_\mu^\mrout (x)) \mathcal{E}^{\mrin \, \mu}(x)\\ \times e^{- \im (\phi_p^\mrout + \phi_k^\mrin - \phi_l^\mrout)} \mathcal{M}_n(l_p, \ldots).
\end{multline}
The determinants of the vielbeins $|E^{\mrin/\mrout}(s)|$ come from the normalisation of the solution to the wave equation. 

\medskip 

We will first consider attaching the ingoing photon to an ingoing scalar:
\begin{equation}
    \omega\, e Q\int \frac{\d^4 x \, \hat{\d}^4 l |E^\mrin (x)|^{-3/2}}{ (l^2 - m^2 + \im \epsilon)} (P^\mrin_\mu(x) + L^\mrin_\mu(x)) \mathcal{E}^{\mrin \, \mu}(x) e^{- \im (\phi_p^\mrin + \phi_k^\mrin - \phi_l^\mrin)} \mathcal{M}_n(l_p, \ldots).
\end{equation}
For this term, we can repeat a lot of the steps from the previous subsections. Since we have matched boundary conditions, the product $(P^\mrin + L^\mrin)\cdot \mathcal{E}^{\mrin}$ only depends on $x^-$~\cite{Adamo:2017nia,Adamo:2020qru}. The exponenent is proportional to (neglecting in labels for the time being)
\begin{multline}
    \phi_k(x) + \phi_p(x) - \phi_l(x) = \frac{k_+ + p_+ - l_+}{2} \sigma_{ab}(x^-) x^a x^b + (k_i + p_i - l_i) E^i_a(x^-) x^a\\
    + (k_+ + p_+ - l_+) x^+ + \Bigg( \frac{p_i p_j}{2p_+} + \frac{k_i k_j}{2k_+} - \frac{l_i l_j}{2l_+}\Bigg) \int^{x^-} \gamma^{ij}(s)\,  \d s + \frac{m^2 \, x^-}{2p_+}- \frac{l^2 \, x^-}{2l_+}.
\end{multline}
Evaluating the $x_+, x_{\perp}$ localises the propagator momentum components 
\begin{align}
    l_+ &= p_+ + k_+, \\
    l_\perp &= p_\perp + k_\perp,
\end{align}
where the integral over $x^\perp$ comes along with a Jacobian factor $|E(x^-)|$. To leading order in the soft momentum we are left with 
\begin{multline}
    2 \omega\, e Q \int \frac{\d x^- \, \hat{\d} (l^2)}{2p_+ |E^{\mrin}(x^-)|^{1/2} (l^2 - m^2 + \im \epsilon)} P^{\mrin}_{\mu} \mathcal{E}^{\mrin \, \mu}(x^-) \exp \Bigg[ \frac{\im (l^2 - m^2)}{2l_+} (x^- - y^-)\Bigg] \\
    \times \exp \Bigg[- \im  \omega \Big( \frac{\hat{k}_i \hat{k}_j}{2 \hat{k}_+} - \frac{\hat{k}_i p_j}{2p_+} + \frac{\hat{k}_+ p_i p_j}{2p_+^2}) \int^{x^-} \gamma^{ij}(s) \,\d s  - \im \omega \frac{\hat{k_+} m^2}{2p_+}  \Bigg] \mathcal{M}_n^y(l, \ldots). 
\end{multline}
Evaluating the integral over $l^2$ in the usual way, and remembering that $p_+ >0$ for this ingoing scalar, we have 
\begin{multline}
    - \frac{\im \omega }{p_+} \int^{y^-}_{-\infty} \frac{\d x^-}{|E(x^-)|^{1/2}} P^{\mrin}_\mu \mathcal{E}^{\mrin \, \mu} (x^-) \exp \Bigg[ \frac{\epsilon(x^- - y^-)}{2p_+} \Bigg] \\
    \times \exp \Bigg[- \im \omega \Big( \frac{\hat{k}_i \hat{k}_j}{2 \hat{k}_+} - \frac{\hat{k}_i p_j}{2p_+} + \frac{\hat{k}_+ p_i p_j}{2p_+^2}) \int^{x^-} \gamma^{ij}(s) \,\d s - \im \omega \frac{\hat{k_+} m^2}{2p_+} \Bigg] \mathcal{M}_n^y(l, \ldots). 
\end{multline}
This is the same type of integral we were considering in both electromagnetism and Yang-Mills and we can use the result from Section \ref{app:func1} to evaluate this in the limit $\omega \rightarrow 0$. All dependence on $y^-$ in the prefactor drops out and we are left with 
\begin{equation}
    \frac{\epsilon \cdot p}{\hat{k} \cdot p} \times \mathcal{M}_n(p, \ldots)
\end{equation}
when the ingoing photon attached directly to an outgoing scalar. 

\medskip 

Let us now consider an ingoing photon attaching directly to an external outgoing scalar with $p_+ < 0$. We consider the term
\begin{equation}
    \omega\, e Q \int \frac{\d^4 x \, \hat{\d}^4 l\, |E^\mrout(x)|^{-1}}{ |E^\mrin(x)|^{1/2} (l^2 - m^2 + \im \epsilon)} (P_\mu^\mrout (x) + L_\mu^\mrout (x)) \mathcal{E}^{\mrin \, \mu}(x) e^{- \im (\phi_p^\mrout + \phi_k^\mrin - \phi_l^\mrout)} \mathcal{M}_n(l_p, \ldots).
\end{equation}
Differences from the previous calculations already enter at the contraction between momentum and polarisation due to the mismatched asymptotics 
\begin{multline}
    (P_\mu^\mrout (x) + L_\mu^\mrout(x)) \mathcal{E}^{\mrin \, \mu}(x) = \epsilon^a \Bigg( \frac{p_+ + l_+}{k_+} k_j E^{\mrin \, j}_a - (p_j + l_j) E^{\mrout \, j}_a \Bigg) \\ + \epsilon^a (p_+ + l_+) (\sigma_{ab}^\mrin - \sigma_{ab}^\mrout) x^b. \label{sec2:PEps}
\end{multline}
Notably this has contributions linear in $x_\perp$. However, this can be cured by inspecting the exponenent in the integral, which is proportional to 
\begin{multline}
    \phi_p^\mrout + \phi_k^\mrin - \phi_l^\mrout = \frac{k_+}{2} x^a x^b (\sigma^\mrin_{ab} - \sigma^\mrout_{ab}) + x^a (k_i E^{\mrin \, i}_a + (p_i - l_i) E^{\mrout \, i}_a) \\
    + (k_+ + p_+ - l_+) x^+ + \Bigg( \frac{p_i p_j}{2p_+} - \frac{l_i l_j}{2l_+} \Bigg) F^{\mrout \, ij} + \frac{k_i k_j}{2k_+} F^{\mrout\, ij} + \frac{m^2}{2p_+}x^- - \frac{l^2}{2l_+} x^-
\end{multline}
on the support of the $\delta$-function in the $x^+$ direction localising $l_+ = p_+ + k_+$. Notice that the quadratic-in-$x^\perp$ part of this exponent is proportional to the linear part of \eqref{sec2:PEps}. This means that we can remove the explicit $x^\perp$ dependence using integration by parts, and neglecting boundary contributions, the original integral is equal to 
\begin{multline}
    \omega\, e Q\int \frac{\d^4 x \, \hat{\d}^4 l\, |E^\mrout (x)|^{-1} }{|E^\mrin(x)|^{1/2}  (l^2- m^2 + \im \epsilon)} \,\epsilon^a E^{\mrout \, j}_a (x^-) \Bigg( (p_j + l_j) - \frac{p_+ + l_+}{k_+} (p_j - l_j)  \Bigg) \\
    \times e^{- \im (\phi_p^\mrout + \phi_k^\mrin - \phi_l^\mrout)} \mathcal{M}_n(l_p, \ldots). 
\end{multline}
The $x^+$ integral fixes $l_+ = p_+ + k_+$, whereas the $x^\perp$ is a Gaussian which integrates to
\begin{multline}
    \omega\, e Q \int \frac{\d x^- \, \hat{\d}^3 l \, |E^\mrout (x)|^{-1} }{|E^\mrin(x)|^{1/2} (l^2 - m^2+ \im \epsilon)}\, \epsilon^a E^{\mrout \, j}_a (x^-) \Bigg( (p_j + l_j) - \frac{p_+ + l_+}{k_+} (p_j - l_j) \Bigg) \\ 
    \times \frac{2 \pi}{k_+ \sqrt{\det (A)}} \exp \Big[ - \frac{\im}{2k_+} B^T A^{-1} B - \im C\Big] \label{sec2:gravGaussInt}
\end{multline}
with definitions 
\begin{align}
    A_{ab} &\coloneqq \sigma_{ab}^\mrin(x^-) - \sigma_{ab}^\mrout (x^-), \label{photAdef} \\
    B_a &\coloneqq E^{\mrout \, i}_a (p_i - l_i) + E^{\mrin \, i}_a k_i,  \label{photBdef}\\
    \tilde{C} & \coloneqq \Bigg(\frac{p_i p_j}{2p_+} - \frac{l_i l_j}{2l_+} \Bigg) F^{ij}_\mrout + \frac{k_i k_j}{2k_+} F^{ij}_\mrin + \frac{m^2 x^-}{2p_+} - \frac{l^2 x^-}{2l_+}. \label{photCdef}
\end{align}
Since we're taking the soft limit, the $l_\perp$ integrals can be localised with help of stationary phase on the exponent in \eqref{sec2:gravGaussInt} as $\omega \rightarrow 0$. This localises onto solutions of $B_a = 0$:
\begin{equation}
    E^{ \mrout\, i}_a l_i = E^{ \mrout\, i}_a p_i + E^{\mrin \, i}_a k_i.
\end{equation}
Notice that to leading order in $\omega$ this implies that $l_\perp = p_\perp$ for fixed $x^-$. The Jacobian from localising cancels some of the determinants and we have the penultimate integral, leading in $\omega$: 
\begin{multline}
    \omega\, e Q \int \frac{\d x^- \, \d (l^2)\, |E^\mrin(x^-)|^{-1/2}}{p_+ \,  (l^2 - m^2+ \im \epsilon)}\, \epsilon^a E^{\mrout \, j}_a \Big(p_j + \frac{p_+}{k_+} k_i E^{\mrin \, i}_b E^{\mrout \, b}_j \Big) e^{- \im  \omega C(x^-) }   \\
    \times e^{\im (l^2 - m^2) (x^- - y^-)/2p_+} \, \mathcal{M}_n^y(l, \ldots)
\end{multline}
where 
\begin{equation}
    C(x^-) = \Bigg( \frac{p_i p_j}{2p_+} \hat{k}_+ - \frac{p_i E^\mrout_{j\, a} E^{\mrin \, l  a} \hat{k}_l}{p_+}\Bigg) F^{ij}_\mrout + \frac{\hat{k}_i \hat{k}_j}{2\hat{k}_+} F^{ij}_\mrin + \frac{\hat{k}_+ m^2 x^-}{2p_+}
\end{equation}
is $\tilde{C}$ evaluated on the solutions $B_a = 0$, subtracting $(l^2 - m^2)/x^-/2p_+$ and rescaled. Finally doing the $l^2$ integral, remembering that $p_+ < 0$, we have 
\begin{multline}
    \omega\, e Q \int_{y^-}^\infty \frac{\d x^-}{p_+ |E^\mrin(x^-)|^{1/2}} \,\epsilon^a E^{\mrout \, j}_a \Big( p_j + \frac{p_+}{k_+} k_i E^{\mrin \, i}_b E^{\mrout \, b}_j\Big) e^{- \im \omega  C(x^-)}\,  e^{\epsilon(x^- - y^-)/2p_+} \\
    \times \mathcal{M}_n^y(l, \ldots). \label{sec2:photonLast}
\end{multline}
Taking the limit $\epsilon \rightarrow 0$ at this point, this is an integral of the form 
\begin{equation}
    \omega \int_{y^-}^\infty f(x^-) e^{\im \omega g(x^-)}\,  \d x^-,
\end{equation}
where the asymptotic behaviours of the functions $f, g$ as $x^- \rightarrow \infty$ are 
\begin{equation}
    f(x^-) \sim 1/x^- , \quad g(x^-) \sim x^- \quad \text{as } x^- \rightarrow \infty.
\end{equation}
This can be checked by refering to Table \ref{sec1:scaleTable}. We show in Appendix \ref{app:func2} that 
\begin{equation}
    \lim_{\omega \rightarrow 0}  \omega \int_{y^-}^\infty f(x^-) e^{\im \omega g(x^-)}\,  \d x^-  = 0.
\end{equation}
Therefore, attaching an ingoing soft photon to an outgoing scalar leg in a gravitational background does not contribute to the soft singularity. Of note here is that this fall-off is only possible to obtain assuming that $|E^\mrin| \sim 1/x^-$. Specifically, this is \emph{not} the case for a self-dual or perturbative plane wave. We will consider that case separately in Section \ref{sec:Comp}.

The same analysis can be repeated with an outgoing soft photon, in which case we obtain the usual soft terms when attaching to an \emph{outgoing} scalar, and no soft singularity when attaching to an \emph{ingoing} scalar.

\medskip 

The leading soft photon theorem on a gravitational plane wave with velocity memory therefore has two variants.
\begin{tcolorbox}[colback=red!5!white,colframe=red!75!black,title=Soft photon theorem on a gravitational plane wave background]\vspace{-0.35cm}
    \begin{gather}
    \lim_{\omega \rightarrow 0} \omega \mathcal{M}_{n+1}(\{p_i\}; \omega \hat{k}^\mrin) = e \sum_{p_i \text{ ingoing}}Q_i \,  \frac{\epsilon \cdot p_i}{\hat{k} \cdot p_i}\, \mathcal{M}_n(\{p_i\}), \label{photonGR1}\\
    \lim_{\omega \rightarrow 0} \omega \mathcal{M}_{n+1} (\{p\}, \omega \hat{k}^\mrout) = e \sum_{p_i \text{ outgoing}} Q_i \,\frac{\epsilon \cdot p_i}{\hat{k} \cdot p_i} \, \mathcal{M}_n(\{p_i\}).\label{photonGR2}
\end{gather}
\end{tcolorbox}
The lack of coupling to particles with opposing boundary conditions can be seen as a consequence of the soft photon not having a definite momentum in the region opposite to where it's defined (see Figure~\ref{fig:photonGR}), as well as the particular scaling behaviour of $E_{i\, a}(x^-)$ and other background quantities. This scaling is crucially due to considering a plane wave spacetime with velocity memory. 

\subsubsection{Soft graviton on gravitational plane waves} \label{sec2:grav}

\begin{figure}
    \centering
    \includegraphics[width=0.55\textwidth]{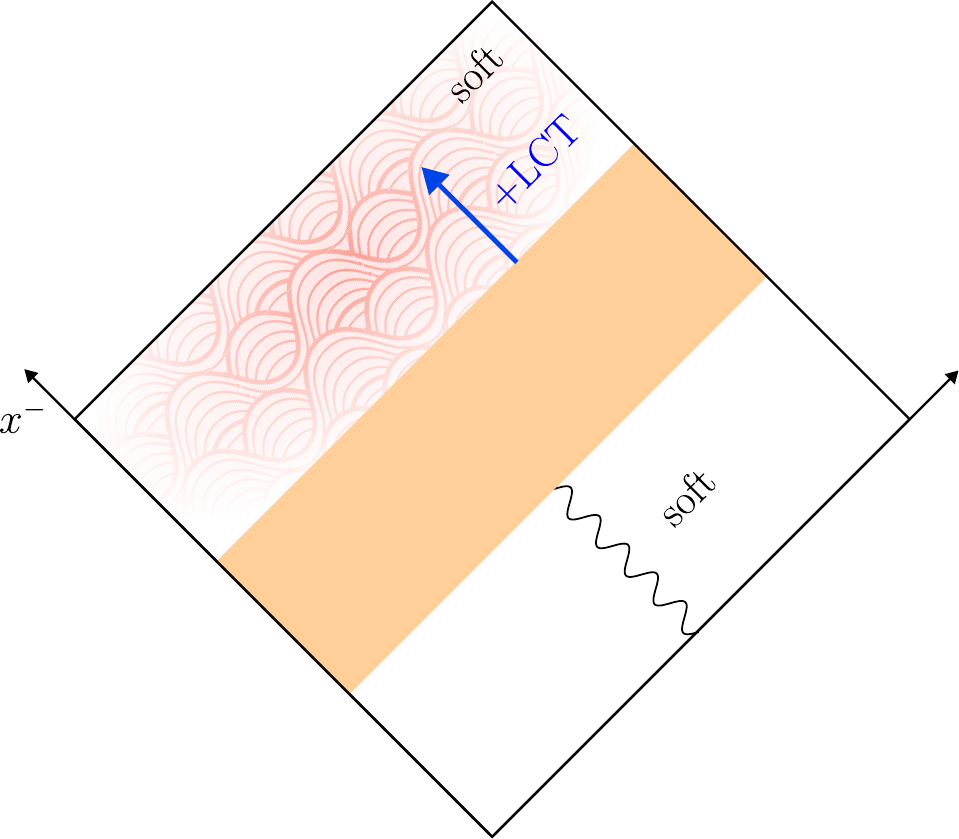}
    \caption{A graviton defined on a gravitational plane wave that is soft in the ingoing region will `spread' out in the outregion and introduce a large coordinate transformation of the background metric.} \label{fig:graviton}
\end{figure}

Just like gluons on a gauge theory background, the polarisation tensor for gravitons is infrared divergent:
\begin{equation}
    \mathcal{E}^{\mu \nu} (x^-)= \mathcal{E}^\mu (x^-) \mathcal{E}^{\nu}(x^-) - \frac{\im}{k_+} \epsilon_a \epsilon_b \sigma^{ab}(x^-)n^{\mu} n^\nu. 
\end{equation}
Assuming $\sigma_{ab}\epsilon^a \epsilon^b \neq 0$, this causes a soft divergence whenever the external soft graviton couples to anything including internal propagators and vertices. But in contrast to Yang-Mills, the momentum of the graviton stays soft throughtout the scattering process. This means that this term is not naturally suppressed using the rapidly oscillating integrals we saw earlier. However, we can view this term as just part of the background metric. Taking the soft limit directly 
\begin{equation}
      \lim_{\omega \rightarrow 0} - \omega \frac{ \im \, \kappa }{k_+} \epsilon_a \epsilon_b \sigma^{ab}(x^-) n^\mu n^\nu \Phi(x) = - \frac{\im \, \kappa}{\hat{k}_+ \sqrt{|E(x^-)|}} \epsilon_a \epsilon_b \sigma^{ab} (x^-) n_{\nu}n_{\mu}. \label{sec2:divEps}
\end{equation}
Notably this only depends on $x^-$. Viewing this metric perturbation as a part of the background we have
\begin{equation}
    \d s^2 \rightarrow \d s^2 - \frac{\im \, \kappa}{k_+ \sqrt{|E(x^-)|}} \epsilon_a \epsilon_b \sigma^{ab}(x^-) (\d x^-)^2,
\end{equation}
which is a pp-wave metric. But in fact, applying a large coordinate transformation to the $x^+$ coordiantes 
\begin{equation}
    x^+ \rightarrow x^+ + \kappa \, f(x^-), \quad f(x^-) =\int^{x^-}  \frac{\im}{2k_+ \sqrt{|E(x^-)|}} \epsilon_a \epsilon_b \sigma^{ab}(s) \, \d s 
\end{equation}
this is equivalent to the original metric. The reason this is a `large' coordinate transformation is that generally (assuming ingoing boundary conditions so that we take the $x^- \rightarrow \infty$ limit) 
\begin{equation}
    f(\infty) \coloneqq \lim_{x^- \rightarrow \infty} f(x^-) \neq 0.
\end{equation}
The effect of this large diffeomorphism on the scattering amplitude is to dress all of our external states by an additional phase~\cite{Cristofoli:2022phh}. The new scalar solutions are 
\begin{equation}
    \Phi_p (x^+) \rightarrow \Phi_p(x^+ + f(x^-)) = e^{\im \, \kappa \, p_+ f(x^-)}\Phi_p (x). 
\end{equation}
It turns out that the amplitudes only see this phase in the LSZ truncation. Since the spacetime is physically equivalent to the original spacetime, the rest of the scattering process will not be affected. If we are attaching an ingoing graviton, the effect on the states in the asymptotic past/future is 
\begin{gather}
    \lim_{x^- \rightarrow -\infty} e^{\im \kappa p_+ f(x^-)} \Phi^{\mrin}_p (x) = \Phi^{\mrin}_p(x), \\
    \lim_{x^- \rightarrow + \infty} e^{\im \kappa p_+ f(x^-)} \Phi^{\mrout}_p (x) = e^{\im \kappa p_+ f(\infty)} \Phi_p^\mrout (x).
\end{gather}
In order to truncate on the correct external states, we must compensate by the factor 
\begin{equation}
    \exp \Bigg[ \im\, \kappa \,  f(\infty) \sum_{p_i \text{ outgoing}} p_{i \, +} \Bigg]  = \exp \Bigg[ - \im\, \kappa \,  f(\infty) \sum_{p_i \text{ ingoing}} p_{i \, +} \Bigg],
\end{equation}
where the equivalence of this factor comes from momentum conservation in the $+$-direction. 
Therefore, to leading order in $\kappa$, this part of the ingoing soft graviton contributes an overall factor 
\begin{equation}
    \Bigg[- \frac{\kappa }{2\hat{k}_+} \int_{- \infty}^\infty \frac{\epsilon_a \epsilon_b \sigma^{\mrin \, ab}(s)}{\sqrt{|E^\mrin(s)|}} \d s  \sum_{p_i \text{ ingoing}} p_{i \, +}\Bigg]\times \mathcal{M}_n (\{p_i\}) \label{sec2:tailDiv}
\end{equation}
to the soft limit. One way to view this contribution is as the soft graviton backreacting on the background metric. Because of the strength of the background, this causes a permanent non-linear change in the coordinates.

\medskip 

A soft graviton in a gravitational plane wave therefore has a part that shifts the background by a coordinate transformation, and something that looks like a usual propagating soft graviton in the asymptotic regions given by 
\begin{equation}
    \mathcal{E}^\mu(x) \mathcal{E}^\nu(x) \, \Phi (x). \label{sec2:softgravLO}
\end{equation}
Because of the transverse $x^\perp$ dependence in this term, we need to be a bit careful about where singularities can arise. We will first consider the terms coming from an ingoing soft graviton attaching directly to an ingoing external scalar. Next, we will see what happens when this part of the soft graviton attaches to other parts of the scattering process, including external outgoing scalars. 

\medskip 

For the ingoing soft graviton part \eqref{sec2:softgravLO} to an ingoing scalar we want to look at the term 
\begin{equation}
    \omega \, \frac{\kappa}{2} \int \frac{\d^4 x \, \hat{\d}^4 l\, |E^\mrin (x)|^{- 3/2} }{ (l^2 - m^2+ \im \epsilon)} \Big[(P^\mrin_\mu(x) + L^\mrin_\mu(x)) \mathcal{E}^{\mrin \, \mu}(x)\Big]^2 e^{- \im (\phi_p^\mrin + \phi_k^\mrin - \phi_l^\mrin)} \mathcal{M}_n(l_p, \ldots) 
\end{equation}
in the limit $\omega \rightarrow 0$. This calculation follows precisely the same steps as for the soft photon in Section \ref{sec:photonGR} and we end up with the $\omega \rightarrow 0$ limit 
\begin{equation}
    \frac{(\epsilon \cdot p)^2}{\hat{k} \cdot p} \times \mathcal{M}_n(\{p_i\}) 
\end{equation}
for each ingoing leg with momentum $p$. 

The novelty arises when we have mismatched asymptotics between the scalar and the graviton. Consider an ingoing soft graviton attaching to an outgoing scalar 
\begin{equation}
    \omega \, \frac{\kappa}{2} \int \frac{\d^4 x \, \hat{\d}^4 l\, |E^\mrout (x)|^{-1} }{|E^\mrin (x)|^{1/2} (l^2 - m^2 + \im \epsilon)} \Big[(P^\mrout_\mu(x) + L^\mrout_\mu(x)) \mathcal{E}^{\mrin \, \mu}(x)\Big]^2 e^{- \im (\phi_p^\mrin + \phi_k^\mrin - \phi_l^\mrin)} \mathcal{M}_n(l_p, \ldots).
\end{equation}
Expressing the prefactor as a derivative operator 
\begin{multline}
    (P_\mu^\mrout (x) + L^\mrout_\mu(x)) \mathcal{E}^{\mrin \, \mu} (x)  = \epsilon^a E^{\mrout \, j}_a \Bigg( (p_i + l_i) - \frac{p_+ + l_+}{k_+} (p_j - l_j)\Bigg) \\+ \frac{\im}{k_+} (p_+ + l_+) \epsilon^a \frac{\partial}{\partial x^a}
\end{multline}
there is a remainder term when the second derivative operator doesn't hit the phase. Neglecting the boundary terms (by continuing the $l$ momentum slightly into the complex plane, for example), we have 
\begin{multline}
    \omega \, \frac{\kappa}{2}  \int \frac{\d^4 x \,\hat{d}^4 l}{|E^\mrout(x)| |E^{\mrin}(x)|^{1/2} (l^2 + \im \epsilon)} \Bigg[ \Bigg(\epsilon^a E^{\mrout \, j}_a \Big( (p_j + l_j) - \frac{p_+ + l_+}{k_+} (p_j - l_j)\Big) \Bigg)^2
    \\ -\frac{\im}{k_+} (p_+ + l_+)^2\epsilon^a \epsilon^b (\sigma_{ab}^\mrin (x^-)- \sigma_{ab}^\mrout (x^-)) \Bigg] e^{- \im (\phi_p + \phi_k - \phi_l)} \times \mathcal{M}_n (l_p , \ldots). \label{gravcalcpause}
\end{multline}
From the soft photon calculation, we know that terms that are constant or linear in $l_j$ in the above expression will be subleading according to the scaling arguments. We are then left with the two parts that are different from the soft photon calculation. Firstly, there is now a quadratic dependence on $l_{\perp}$ in the prefactors. Secondly, we have a term proportional to $(\sigma^{\mrin}_{ab}(x^-) - \sigma_{ab}^{\mrout}(x^-))$. Looking more closely at this combination 
\begin{multline}
    \omega \, \frac{\kappa}{2}  \int \frac{\d^4 x \,\hat{d}^4 l}{|E^\mrout(x)| |E^{\mrin}(x)|^{1/2} (l^2 + \im \epsilon)} \Bigg[ \Bigg(\epsilon^a E^{\mrout \, j}_a \, \frac{p_+ + l_+}{k_+} (p_j - l_j) \Bigg)^2\\
    -\frac{\im}{k_+} (p_+ + l_+)^2\epsilon^a \epsilon^b (\sigma_{ab}^\mrin (x^-)- \sigma_{ab}^\mrout (x^-)) \Bigg]  \times  e^{- \im (\phi_p + \phi_k - \phi_l)} \times \mathcal{M}_n (l_p , \ldots).
\end{multline}
From here, we can do the $x^+$ integral, imposing momentum conservation in that direction. Since we've removed the $x_{\perp}$ dependence from the integrand, we can also do the Gaussian integral in $x_{\perp}$ straightforwardly. At the same time, doing the $l^2$ integral using the Feynman prescription imposes the usual ordering on $x^-$ and the first vertex lightfront position $y^-$ in the rest of the amplitude. After this set of integrations we have 
\begin{multline}
     \omega \, \frac{\kappa}{2} \int_{y^-}^\infty \frac{\d x^- \, \hat{\d}^2 l_{\perp}}{2l_+ \, |E^{\mrin}(x)|^{1/2} |E^{\mrout}(x)| } \\ \times \Bigg[ \Bigg(\epsilon^a E^{\mrout \, j}_a \, \frac{p_+ + l_+}{k_+} (p_j - l_j) \Bigg)^2
    -\frac{\im}{k_+}(p_+ + l_+)^2\epsilon^a \epsilon^b (\sigma_{ab}^\mrin (x^-)- \sigma_{ab}^\mrout (x^-)) \Bigg]\\  \times \frac{2 \pi}{k_+ \sqrt{\det (A)}} \exp \Bigg[ - \frac{\im }{2k_+} B^T A^{-1} B - \im C\Bigg]  
\end{multline}
with the definitions, repeated from the photon case in (\ref{photAdef}, \ref{photBdef}, \ref{photCdef}),
\begin{align}
    A_{ab} &\coloneqq \sigma_{ab}^\mrin(x^-) - \sigma_{ab}^\mrout (x^-), \\
    B_a &\coloneqq E^{\mrout \, i}_a (p_i - l_i) + E^{\mrin \, i}_a k_i, \\
    \tilde{C} & \coloneqq \Bigg(\frac{p_i p_j}{2p_+} - \frac{l_i l_j}{2l_+} \Bigg) F^{ij}_\mrout + \frac{k_i k_j}{2k_+} F^{ij}_\mrin + \frac{m^2 x^-}{2p_+} - \frac{l^2 x^-}{2l_+}.
\end{align}
Now, define the variable
\begin{equation}
    L_a \coloneqq E_a^{\mrout \, j} (p_j - l_j).
\end{equation}
To leading order in $\omega$, the integral can be rewritten as 
\begin{multline}
     \omega \, \frac{\kappa}{2} \int_{y^-}^\infty \frac{\d x^- \, \hat{\d}^2 L_{\perp}}{2l_+ \, |E^{\mrin}(x)|^{1/2} } \\ \times \Bigg[\Bigg( (p_+ + l_+)^2 \Bigg( \Big(\epsilon_a \frac{\partial}{\partial L_a} \Big)^2 +\frac{\im}{k_+} \epsilon_a \epsilon_b A^{ab} \Bigg)
    -\frac{\im}{k_+}(p_+ + l_+)^2\epsilon^a \epsilon^b (\sigma_{ab}^\mrin (x^-)- \sigma_{ab}^\mrout (x^-)) \Bigg]\\  \times \frac{2 \pi}{k_+ \sqrt{\det (A)}} \exp \Bigg[ - \frac{\im }{2k_+} B^T A^{-1} B - \im C\Bigg] 
\end{multline}
where the new term in the first set of brackets comes again from the quadratic part of the differential operator. This cancels against the second term, and we find that this is once again just a boundary contribution. This entire contribution is therefore subleading in the soft limit. Going back to the original expression~\eqref{gravcalcpause}, we see that all remaining terms will look like those in the derivation of the soft photon theorem on a gravitational plane wave. The final integral will have the form similar to~\eqref{sec2:photonLast}
\begin{multline}
    \omega\, \frac{\kappa}{2} \int_{y^-}^\infty \frac{\d x^-}{p_+ |E^\mrin(x^-)|^{1/2}} \, \Bigg[\epsilon^a E^{\mrout \, j}_a \Big( p_j + \frac{p_+}{k_+} k_i E^{\mrin \, i}_b E^{\mrout \, b}_j\Big)\Bigg]^2 e^{- \im \omega  C(x^-)}\,  e^{\epsilon(x^- - y^-)/2p_+} \\
    \times \mathcal{M}_n^y(l, \ldots). \label{gravFinal}
\end{multline}
Using the same scaling arguments as in that section, we see that this contribution is subleading in the soft limit $\omega \rightarrow 0$. Therefore, the $\mathcal{E}_{\mu}(x)\mathcal{E}_{\nu}(x) \, \Phi(x)$ part of the ingoing graviton wavefunction contributes in the soft limit only when coupling to other ingoing particles. 

\medskip

Another source of possibly divergent contributions from the quadratic-in-$x^{\perp}$ terms when attaching the soft graviton onto internal lines or higher point vertices. It is possible to show because there is no soft phase in these cases, the $\sigma_{ab}^\mrin - \sigma_{ab}^\mrout$ term will not have the $1/k_+$ prefactor as was the case considered above. Therefore these terms will have no divergent behaviour to begin with. The only divergence from attaching to internal propagators would have come from the explicit tail term which has already been addressed above via a large coordinate transformation.

\medskip 

The leading soft graviton theorem on a gravitational plane wave with velocity memory has two variants.
\begin{tcolorbox}[colback=red!5!white,colframe=red!75!black,title=Soft graviton theorem on a gravitational plane wave background]\vspace{-0.35cm}
    \begin{multline}
    \lim_{\omega \rightarrow 0} \omega \mathcal{M}_{n+1}(\{p_i\}; \omega \hat{k}^\mrin)  \\  = \frac{\kappa}{2} \sum_{p_i \text{ ingoing}} \, \Bigg[ \frac{(\epsilon \cdot p_i)^2}{\hat{k} \cdot p_i}-   \frac{p_{i \, +}}{\hat{k}_+} \, \epsilon^a \epsilon^b \,  \int_{-\infty}^{\infty}\frac{\sigma^\mrin_{ab}(s)}{|E^{\mrin}(s)|^{1/2}} \d s\Bigg]\, \mathcal{M}_n(\{p_i\}), \label{gravitonGR11}\end{multline}

    \begin{multline}\lim_{\omega \rightarrow 0} \omega \mathcal{M}_{n+1} (\{p\}, \omega \hat{k}^\mrout) \\= \frac{\kappa}{2} \sum_{p_i \text{ outgoing}}  \,\Bigg[ \frac{(\epsilon \cdot p_i)^2}{\hat{k} \cdot p_i} -  \frac{p_{i \, +}}{\hat{k}_+} \, \epsilon^a \epsilon^b \,  \int_{-\infty}^{\infty}\frac{\sigma^\mrout_{ab}(s)}{|E^{\mrout}(s)|^{1/2}} \d s\Bigg]\, \mathcal{M}_n(\{p_i\}),\label{gravitonGR22}
\end{multline}
where $\sigma^{\mrin/\mrout}_{ab}$ and $E^{\mrin/\mrout}_{i\, a}$ are defined in (\ref{sec1:deform}, \ref{sec1:VielEq}) respectively.
\end{tcolorbox}
The extra term propotional to $\sigma_{ab}$ can be associated with the tail effect, present in gravitational radiation on a plane wave background. In the soft limit, this generated a large coordinate transformation.

\subsection{Comments on gauge invariance}

One aspect that is striking about the soft gluon theorem on a gauge theory background (\ref{gluon1}, \ref{gluon2}) and the soft photon and graviton theorems on a gravitational background (\ref{photonGR1}, \ref{photonGR2}) is that they only contain half of the terms appearing in the flat space results. This is particularly suprising when one recalls that gauge invariance of the flat space result implies charge conservation, a fact which requires summation overl \emph{all} external particles. For example, when we consider the soft photon theorem on flat space
\begin{equation}
    \lim_{\omega \rightarrow 0} \omega \mathcal{M}_{n+1}(\{p_i\}; \omega \hat{k}_\text{photon})  = \sum_{p_i \text{ external}} Q_i \, \frac{\epsilon \cdot p_i}{\hat{k} \cdot p_i} \, \mathcal{M}_n(\{p_i\}),
\end{equation}
checking for gauge invariance corresponds to replacing $\epsilon \rightarrow A k$ with $A$ some constant. Any amplitude under this redefinition should evaluate to zero. In the above expression, we get zero provided that 
\begin{equation}
  \sum_{p_i \text{ external}} Q_i = 0,
\end{equation}  
i.e. charge is conserved. The statement in gauge theory is more complicated to express due to action of the gauge group and colour orderings but we find that the flat space soft theorem becomes
\begin{equation}
    \Bigg[\sum_{p_i \text{  ingoing}} T^a_i + \sum_{p_i \text{ outgoing}} (T^a_i)^{\ast} \Bigg]\mathcal{M}_n(\{p_i\}) = 0 
\end{equation}
where $T^a_i$ and $(T^a_i)^{\ast}$ act on the $i^\text{th}$ index of $\mathcal{M}$. This quantity is zero as a consequence of the Ward identity for gauge transformation in Yang-Mills. For completeness, in gravity, the statement of gauge invariance becomes 
\begin{equation}
    \Bigg[\sum_{p_i \text{ external}} \epsilon \cdot p_i\Bigg] \mathcal{M}_n(\{p_i\}) = 0
\end{equation}
which is equivalent to momentum conservation with our conventions for the external momenta. 

\medskip 
In contrast, naively  replacing $\epsilon \rightarrow k$ in the soft photon theorem on a gravitational plane wave seems to require 
\begin{equation}
    \sum_{p_i \text{ ingoing}} Q_i \,\mathcal{M}_n(\{p_i\}) = 0
\end{equation}
for gauge invariance, which is of course not the case generally. The problem with checking for `gauge invariance' in this way on a background is three-fold. Firstly, we have already fixed all gauge degrees of freedom in choosing (appropriately) transverse, traceless and lightfront gauges for the linearised perturbations we're considering. This means that there are no additional degrees of freedom one can check for, and it makes no sense to check the amplitude for `gauge invariance'. The only way it would make sense to do so is to consider a dressed polarisation that is not completely gauge fixed, but since the gauge fixing is what allowed us to solve the equations of motion so succinctly, leaving this degree of freedom would severely complicate the calculations. 

\medskip 

But, treating gauge transformations more serious, on a background field a gauge transformation would be generally described by transforming the polarisation tensor to
\begin{equation}
    \mathcal{E}_{\mu}(x) \Phi(x) \rightarrow \mathcal{E}_{\mu}(x) \Phi(x) + \nabla_{\mu} (f(k)\, \Phi_k(x))
\end{equation}
for an arbitrary function of momentum $f(k)$. The dressed polarisation tensor transforms to a function with spatial dependence, rather than being a pure replacement rule. This spatial dependence needs to be accounted for in all integrals, and is expected to vanish as boundary terms in the final results, following from the gauge invariance of the action itself. Gauge invariance of amplitudes in gauge theory plane wave backgrounds has also been explored in~\cite{Ilderton:2020rgk}.

\medskip

Further, it can be argued that checking for gauge invariance under reasonable gauge transformations does not even enter at this order in the energy scale $\omega$. Consider the ingoing soft photon theorem on a gravitational plane wave background \eqref{photonGR1}:
\begin{equation}
\lim_{\omega \rightarrow 0} \omega \mathcal{M}_{n+1}(\{p_i\}; \omega \hat{k}^\mrin) = e \sum_{p_i \text{ ingoing}}Q_i \,  \frac{\epsilon \cdot p_i}{\hat{k} \cdot p_i}\, \mathcal{M}_n(\{p_i\}). \label{rephotonGR}
\end{equation}
Note that the RHS also has an implicit limit of $\omega \rightarrow 0$ in the expression for the polarisation vector since this statement is only valid for low energies. Now consider a gauge transformation for the polarisation of the soft photon $\epsilon_\mu \rightarrow f(k) k_{\mu}$. The function $f(k)$ needs to behave at least as $\omega^{-1 + \delta}$ as $\omega \rightarrow 0$  with $\delta>0$ for this to correspond to a gauge transformation that preserves the appropriate fall-off conditions at $\scri^{\pm}$ for the field\footnote{After Fourier transforming, this describes a gauge transformation $\phi(x) \sim  r^{-1 - \delta}\bar{\phi}(u, \hat{x})$ as $r \rightarrow \infty$ in Bondi coordinates on $\mathcal{I}^+$. The $\delta>0$ condition ensures that this gauge transformation does not affect the asymptotic data of the gauge field~\cite{Strominger:2017zoo}.}. But then the RHS of \eqref{rephotonGR} is 
\begin{equation}
    e \lim_{\omega \rightarrow 0}\omega f(k) \Bigg[ \sum_{p_i \text{ ingoing}} Q_i   \, \mathcal{M}_n(\{p_i\})\Bigg] = 0. 
\end{equation}
Gauge invariance is therefore a trivial statement at this order in the energy scale. It is quite interesting that flat space soft theorems do not care about this fall-off condition - soft theorems in flat space satisfy this gauge invariance even under large gauge transformations as $\omega \rightarrow 0$ (provided there's charge conservation).

\section{Comparisons with known results} \label{sec:Comp}

In this section we will comment on various aspects and special cases of the soft theorems derived in the previous section, and how to compare these results to the literature and flat space. 

\subsection{On self-dual backgrounds} \label{sec:SD}

In this section we consider the special case of a self-dual background. As mentioned in the previous section, many of the analytic techniques used to derive the soft theorems don't apply when we have a self-dual background. In this section we derive the soft theorems (or prove their non-existence) on self-dual backgrounds in gauge theory and gravity. We will also compare these expressions to the all-multiplicity formulae for MHV scattering amplitudes on self-dual backgrounds~\cite{Adamo:2020syc,Adamo:2020yzi,Adamo:2022mev}.

\subsubsection{Yang-Mills}

In deriving the leading soft gluon theorem in the previous section, we relied on a rapidly oscillating phase in the integral. On a self-dual background, the massless scalar solution to the wave equation is not rapidly oscillating for an initially soft particle:
\begin{equation}
    \Phi(x) = \exp \Bigg[ \im \Big( k_+ x^+ + k_\perp x^\perp + \frac{k_\perp k_\perp}{2k_+} + \frac{e_k}{k_+} \int^{x^-} k_{\perp} A^{+ \, \perp} (s) \, \d s \Big) \Bigg].
\end{equation}
However it is still true that a soft particle doesn't stay soft in the opposing region. One should therefore expect that the soft theorems will still be modified on these backgrounds. 

\paragraph{Positive helicity soft gluon}
A positive helicity gluon has wavefunction 
\begin{equation}
    a_{\mu}^{(+)} (x) = \Big( \epsilon_a^{(+)} \delta^a_\mu + \frac{1}{k_+} (k^a + e \mathsf{A}^a) \epsilon^{(+)}_a n_{\mu}\Big) \Phi(x).
\end{equation}
Since $\epsilon^{(+)}_\perp \mathsf{A}^\perp \neq 0$ this has a $1/\omega$ divergence as $\omega \rightarrow 0$. In the non-chiral case, this divergence was suppressed by the rapidly oscillating phase. But since we do not have a rapidly oscillating phase in this case, this soft singularity is present at every vertex where the gluon is attached. In particular, there is no factorisation in the soft limit. Therefore, there does not exist a soft gluon theorem for positive helicity gluons on self-dual backgrounds. 

This matches the expectations from the formula for the MHV gluon scattering amplitude on a self-dual plane wave~\cite{Adamo:2020syc,Adamo:2020yzi} given by 
\begin{multline}
    \mathcal{M}_n^{\text{MHV}} = g_{\text{YM}}^{n-2} \delta^3_{+, \perp} \Bigg(\sum_{i = 1}^n k_i \Bigg) \frac{ \langle r \, s \rangle^4}{\langle 1 \, 2 \rangle \, \langle 2 \, 3\rangle \, \cdots \, \langle n -1\, n \rangle \langle n \, 1 \rangle}  \\ \times \int_{- \infty}^\infty \d x^- \, \exp \, \Bigg[ \im \sum_{i = 1}^n \int^{x^-} \d s \, K_{i \, -}(s)\Bigg] \label{sec3:GluMHV}
\end{multline}
with $r, s$ labelling the two negative helicity gluons, and this being the colour-ordered amplitude corresponding to the colour-ordering $123 \ldots n$. Here $K_{i \, -}$ are the dressed lightfront  momenta for the particles. On a self-dual background 
\begin{equation}
    K_{i \, -}(s) = \frac{k_\perp k^\perp}{2k_+} + e_i \frac{k_\perp \mathsf{A}^\perp(s)}{k_+}.
\end{equation}
Because $\lim_{\omega \rightarrow 0} K_{i \, -} \neq 0$ whenever $e_i \neq 0$, this amplitude does not factorise in the soft limit for a positive helicity gluon, as expected from the non-existence of a soft theorem shown above.

\paragraph{Negative helicity soft gluon}
The transverse polarisation vector for negative helicity gluon satisfies $\epsilon^{(-)}_\perp \mathsf{A}^\perp = 0$. Therefore a negative helicity gluon has wavefunction 
\begin{equation}
    a_{\mu}^{(-)} (x) = \Big( \epsilon_a^{(-)} \delta^a_\mu + \frac{1}{k_+} k^a \epsilon^{(-)}_a n_{\mu} \Big)\Phi(x).
\end{equation}
This no longer has an inherent $1/\omega$ singularity, which means that the only soft singularities arise when the soft gluon is directly attached to an external particle and a propagator goes on-shell. However, since a soft ingoing gluon is only soft in the in-going region, a simple adaptation of the arguments in Section \ref{sec:YM} means that only attaching to \emph{ingoing} external particles contributes to the leading soft theorem with all other terms being suppressed. Therefore we recover the leading soft negative helicity gluon theorems on a self-dual plane wave in gauge theory.
\begin{tcolorbox}[colback=blue!5!white,colframe=blue!75!black,title=Soft negative helicity gluon theorem on a self-dual gauge theory plane wave]\vspace{-0.35cm}
    \begin{gather}
    \lim_{\omega \rightarrow 0} \omega \mathcal{M}_{n+1}(\{p_i\}; \omega \hat{k}^{\mrin\, (-)}) = g_{\text{YM}} \sum_{p_i \text{ ingoing}} \frac{\epsilon^{(-)} \cdot p_i}{\hat{k} \cdot p_i}  T^\mathsf{a}_{p_i}\mathcal{M}_n (\{p_i\}) \\ 
    \lim_{\omega \rightarrow 0} \omega \mathcal{M}_{n+1}(\{p_i\}; \omega \hat{k}^{\mrout\, (-)}) = g_{\text{YM}}\sum_{p_i \text{ outgoing}} \frac{\epsilon^{(-)} \cdot p_i}{\hat{k} \cdot p_i}  (T^\mathsf{a}_{p_i})^{\ast}\mathcal{M}_n (\{p_i\})
\end{gather}
\end{tcolorbox}

This is also consistent with the MHV amplitude \eqref{sec3:GluMHV}, but only because the amplitude with one negative helicity gluon and rest positive vanishes. It would be interesting to explore whether the N$^{k}$MHV amplitude formula conjectured in~\cite{Adamo:2020yzi} is consistent with these results.

\subsubsection{Gravity}

Whilst self-duality of the background drastically changed a key aspect of the scalar solution in gauge theory, the effect is not as apparently drastic in gravity. The main effect is a change in scalings of the key geometric objects. In Table \ref{sec3:sdTable} we collect the large-$x^-$ behaviour of common quantities in a self-dual gravitational background assuming a sandwich profile in Brinkmann coordinates. This should be contrasted with Table \ref{sec1:scaleTable}. 
In all of the following cases we will see that these scalings play a key role. 
\begin{table}
    \centering
    \begin{tabular}{c || c | c | c}
        & $x^- \rightarrow -\infty$ & $x^- \rightarrow + \infty$ &  \\ \hline
        $E^{\mrin}_{i \, a}$ & $ 1$ & $ c x^-$ & \eqref{SDViels}\\
        $E^{\mrin \, i}_a$ & $ 1$ & $- c x^-$&\\
        $\sigma^{\mrin}_{ab}$ & 0 &$  - c $& \eqref{SDdeform} \\
        $\gamma^{\mrin}_{ij}$ & $1$ & $2 c x^-$ & \eqref{sec1:ERmet} \\
        $\gamma^{\mrin \, ij}$ & $ 1$ & $- 2 c x^-$ & \eqref{SDERmet}\\
        $F^{\mrin\, ij}$ & $ x^-$ & $- 2 c (x^-)^2$ & \\
        $\det{E^{\mrin}}$ & 1 & 1 & \eqref{SDViels}
    \end{tabular}
    \caption{The asymptotic-in-$x^-$ behaviour of useful geometric quantities in a self-dual gravitational plane wave, ignoring the tensor structure of the quantities. Here $c \propto \int_{-\infty}^\infty \dot{f} \d s$, where $f$ is the wave profile, see \eqref{sec1:sdGravProf}. The behaviour for the outgoing gauge are the same, with $+ \infty$ and $-\infty$ swapped.} \label{sec3:sdTable}
\end{table}

\paragraph{Positive helicity soft photon}

As before, the positive helicity polarisation vector has no special vanishing properties when contracted into background-dependent quantities. We therefore follow all same steps as for the non-chiral derivation of the soft photon theorem on gravitational plane waves in Section \ref{sec:photonGR}. For an ingoing soft photon attaching directly to an ingoing scalar, the soft contribution is precisely the same as no scaling behaviour was used in the calculation. However, attaching to an outgoing scalar, the last step we reach is \eqref{sec2:photonLast} which we recall here for specifically a self-dual background:
\begin{equation}
    \omega \, e Q \int_{y^-}^\infty \frac{\d x^-}{p_+ } \,\epsilon^{(+) \, a} E^{\mrout \, j}_a \Big( p_j + \frac{p_+}{k_+} k_i E^{\mrin \, i}_b E^{\mrout \, b}_j\Big) e^{- \im \omega  C(x^-)}\,  e^{\epsilon(x^- - y^-)/2p_+}
    \times \mathcal{M}_n^y(l, \ldots).
\end{equation}
This is an integral of the form 
\begin{equation}
    \omega\, eQ  \int_{y^-}^{\infty} f(x^-) e^{\im \omega g(x^-)} \d x^-
\end{equation}
where the asymptotic behaviours of the functions $f, g$ are 
\begin{equation}
    f(x^-) \sim x^-, \quad g(x^-) \sim (x^-)^2 \quad \text{as } x^- \rightarrow \infty,
\end{equation}
referring to Table \ref{sec3:sdTable}. Notably, this does not satisfy the necessary condition for being suppressed presented in Appendix \ref{app:func2} since 
\begin{equation}
    \frac{f(x^-)}{g'(x^-)} \sim 1 \qquad \text{as } x^- \rightarrow \infty. 
\end{equation}
However, changing our integration variable to $z = (x^-)^2$ and rewriting $f, g$ as functions of $z$, the integral is now of the form 
\begin{equation}
    \omega \, e Q \int_{(y^-)^2}^\infty \tilde{f}(z)\,  e^{-\im \omega \tilde{g}(z)} \d z
\end{equation}
where 
\begin{equation}
    \tilde{f}(z) \sim \frac{ p_+ }{2\hat{k}_+}\, \epsilon^{(+)\, a} \, \hat{k}_i \, c^i_a, \qquad \tilde{g}(z) \sim \frac{\hat{k}_i\hat{k_j}}{2\hat{k}_+}\,  c^{ij} \, z \qquad \text{as } z \rightarrow \infty.
\end{equation}
This behaviour is a variant of the one studied in Appendix \ref{app:func1}. The conclusion after applying those results to this case is that
\begin{equation}
    \lim_{\omega \rightarrow 0} \omega\, e Q  \int_{y^-}^\infty f(x^-) e^{-\im \omega g(x^-)} \d x^- =e Q \,  \frac{ \epsilon^{(+) \, a} \hat{k}_i c^i_a}{ \hat{k}_i \hat{k}_j c^{ij}}.
\end{equation}
This extra term that wasn't present in a non-chiral background can be interpreted as the usual soft term for a photon with polarisation and momentum 
\begin{equation}
    \epsilon_{\mu} \sim \frac{\epsilon^{(+) \, a} \hat{k}_i c^i_a}{2 \hat{k}_+}\, n_{\mu}, \qquad k_{\mu} \sim \frac{\hat{k}_i \hat{k}_j c^{ij}}{2\hat{k}_+} \, n_{\mu}
\end{equation}
contracted with the outgoing scalars, though whether this should be the physical interpretation is still unclear. Finally, we have found that the amplitude factorises and we have the soft positive helicity photon theorem on a self-dual plane wave.
\begin{tcolorbox}[colback=blue!5!white,colframe=blue!75!black,title=Soft positive helicity photon theorem on a self-dual gravitational plane wave]\vspace{-0.35cm}
\begin{gather}
    \lim_{\omega \rightarrow 0} \omega \mathcal{M}_{n+1} (\{p_i\}; \omega \hat{k}^{\mrin \, (+)}) = e \Bigg[\sum_{p_i \text{ ingoing}} \!\!\!Q_i \frac{\epsilon^{(+)} \cdot p_i}{\hat{k} \cdot p_i} +\!\!\!\sum_{p_i \text{ outgoing}}\!\!\! \!Q_i  \frac{ \epsilon^{(+) \, a} \hat{k}_i c^i_a}{ \hat{k}_i \hat{k}_j c^{ij}} \Bigg] \mathcal{M}_n (\{p_i\}), \\
    \lim_{\omega \rightarrow 0} \omega \mathcal{M}_{n+1} (\{p_i\}; \omega \hat{k}^{\mrout \, (+)}) = e \Bigg[ \sum_{p_i \text{ ingoing}}\!\!\! Q_i  \frac{ \epsilon^{(+) \, a} \hat{k}_i c^i_a}{ \hat{k}_i \hat{k}_j c^{ij}} +\!\! \!\sum_{p_i \text{ outgoing}} \!\!\!\!Q_i \frac{\epsilon^{(+)} \cdot p_i}{\hat{k} \cdot p_i} \Bigg]   \mathcal{M}_n (\{p_i\})
\end{gather}
\end{tcolorbox}
There does not as-yet exist an MHV amplitude for scattering in Einstein-Maxwell on self-dual backgrounds to compare to this result. 

\paragraph{Negative helicity soft photon} 
The negative helicity polarisation vector satisfies 
\begin{equation}
    E^i_a \epsilon^{(-)\, a} = \epsilon^{(-)\, a}, \qquad \sigma_{ab} \epsilon^{(-)\, a} = 0. \label{sec3:softVan}
\end{equation}
Consequently, whilst the in-in argument carries through as before, attaching an ingoing gluon to an outgoing scalar \eqref{sec2:photonLast} culminates in the term 
\begin{equation}
    \omega\, eQ \int_{y^-}^\infty \frac{\d x^-}{p_+ } \,\epsilon^{(-) \, a}\Big( p_a + \frac{p_+}{k_+} k_a\Big) e^{- \im \omega  C(x^-)}\,  e^{\epsilon(x^- - y^-)/2p_+}
    \times \mathcal{M}_n^y(l, \ldots).
\end{equation}
Again, we have found an integral of the form 
\begin{equation}
    \omega\, eQ\int_{y^-}^{\infty} f(x^-) e^{\im \omega g(x^-)} \d x^-
\end{equation}
where the asymptotic behaviour is now
\begin{equation}
    f(x^-) \sim (x^-)^0, \quad g(x^-) \sim (x^-)^2 \quad \text{as } x^- \rightarrow \infty. 
\end{equation}
This time we \emph{can} apply the result of Appendix \ref{app:func2} since 
\begin{equation}
    \frac{f(x^-)}{g'(x^-)} \sim (x^-)^{-1} \qquad \text{as } x^- \rightarrow \infty. 
\end{equation}
Therefore, attaching an ingoing negative helicity soft photon to an outgoing scalar is still suppressed. The soft theorem in this case is therefore:
\begin{tcolorbox}[colback=blue!5!white,colframe=blue!75!black,title=Soft negative helicity photon theorem on a self-dual gravitational plane wave]\vspace{-0.35cm}
\begin{gather}
    \lim_{\omega \rightarrow 0} \omega \mathcal{M}_{n+1}(\{p_i\}; \omega \hat{k}^\mrin) = e \sum_{p_i \text{ ingoing}} Q_i \, \frac{\epsilon^{(-)} \cdot p_i}{\hat{k} \cdot p_i}\, \mathcal{M}_n(\{p_i\}), \\
    \lim_{\omega \rightarrow 0} \omega \mathcal{M}_{n+1} (\{p\}, \omega \hat{k}^\mrout) = e \sum_{p_i \text{ outgoing}} Q_i \, \frac{\epsilon^{(-)} \cdot p_i}{\hat{k} \cdot p_i} \, \mathcal{M}_n(\{p_i\}).
\end{gather}
\end{tcolorbox}

\paragraph{Positive helicity soft graviton}

Here, we will only verify that the amplitude with a positive helicity  soft graviton on a self-dual plane wave does not have an infrared singularity worse than $1/\omega$. The reason one may be concerned about this is when looking at the scaling of the large coordinate transformation from the tail term. Recall that for an ingoing graviton, this takes the form 
\begin{equation}
     \frac{\kappa}{2 \hat{k}_+} \int_{- \infty}^{\infty} \frac{\epsilon^{(+)\, a} \epsilon^{(+)\, b}\sigma^\mrin_{ab}(s)}{|E^{\mrin}(s)|^{1/2}} \d s \, \sum_{p_i \text{ outgoing}} p_{i \, +}.
\end{equation}
For a self-dual plane wave, the leading divergent part\footnote{To study this more rigorously, it would be better to look at $\lim_{\omega^2} \mathcal{M}_{n+1}$ and repeat the analysis for these tail terms and the rest of the amplitude. We will leave this to future endeavours and merely sketch the resolution here, based on the workings of the previous section.} of this integral is
\begin{equation}
    -\frac{\kappa}{2 \hat{k}_+}\Bigg[ \int^{\infty}_{x_f^-} c\,\d s\Bigg] \, \sum_{p_i \text{ outgoing}} p_{i \, +}, \label{sec4:ldiv}
\end{equation}
according to Table \ref{sec3:sdTable} and using properties of contractions between self-dual polarisations and these background quantities. This is of course very divergent, but we will show that it cancels exactly with other parts of the amplitude. 

Going back to expression when the ingoing graviton couples to an outgoing scalar in~\eqref{gravFinal}, this time on a self-dual background, 
\begin{multline}
    \omega\, \frac{\kappa}{2} \int_{y^-}^\infty \frac{\d x^-}{p_+} \, \Bigg[\epsilon^{(+)\, a} E^{\mrout \, j}_a \Big( p_j + \frac{p_+}{k_+} k_i E^{\mrin \, i}_b E^{\mrout \, b}_j\Big)\Bigg]^2 e^{- \im \omega  C(x^-)}\,  e^{\epsilon(x^- - y^-)/2p_+} \\
    \times \mathcal{M}_n^y(l, \ldots)
\end{multline}
note that the dominant-in-$x^-$ part of the integrand is the piece quadratic in $E^{\mrin \, i}_b$:
\begin{equation}
    \omega\, \frac{\kappa}{2}  \int_{y^-}^\infty \frac{\d x^-}{p_+} \, \Bigg[\epsilon^{(+)\, a} E^{\mrout \, j}_a \Big(\frac{p_+}{k_+} k_i E^{\mrin \, i}_b E^{\mrout \, b}_j\Big)\Bigg]^2 e^{- \im \omega  C(x^-)}\,  e^{\epsilon(x^- - y^-)/2p_+}
    \times \mathcal{M}_n^y(l, \ldots).
\end{equation}
Using the expressions for large $x^-$ for everthing appearing here, the main part of the integral in the out-region becomes 
\begin{equation}
    \omega \, \frac{\kappa}{2} \int_{x_f^-}^{\infty}p_+     \frac{c G}{\hat{k}_+}\,(x^-)^2\,  e^{- \omega (x^-)^2 G } \, \d x^- 
\end{equation}
where the constant
\begin{equation}
G = \frac{\hat{k}_i \hat{k}_j c^{ij}}{ \hat{k_+}} 
\end{equation}
using properties of the transverse polarisations of positive helicity gravitons and the matrix $c^{ij}$ for self-dual plane waves. Assuming that there's a reasonable prescription for boundary contributions, integrating by parts gives to leading order in $\omega$ 
\begin{equation}
    \frac{\kappa}{2 \hat{k_+}}\Bigg[ \int_{x_f^-}^{\infty}   c \, \Bigg] p_+.
\end{equation}
Summing this over all outgoing legs, we see that this cancels against the leading divergence from the tail term in \eqref{sec4:ldiv}, meaning that there is no $1/\omega^2$ divergence in the soft limit of the amplitude on this background. 

We will leave further analysis of the other contributions from positive helicity soft gravitons to future work. A parallel avenue is looking at soft limits of the MHV amplitudes for graviton scattering presented in~\cite{Adamo:2022mev}, which take the form of modified Hodges' matrices with tail insertions integrated over a lightfront coordinate. These formulae have already been studied in other contexts relating to the infrared, such as holomorphic collinear limits, in~\cite{Adamo:2023zeh}.

\paragraph{Negative helicity soft graviton}

The analysis for negative helicity soft gravitons is much the same as negative helicity soft photons, due to the vanishing of various background-dependent quantities with the polarisation tensor as in \eqref{sec3:softVan}. This means that negative helicity soft gravitons do not have a divergent tail term, and we need only consider the soft divergences when attaching directly to external legs. The in-in contribution proceeds as in the previous section. Attaching the ingoing soft graviton to an outgoing scalar yields a term similar to the photon case 
\begin{equation}
    \omega \, \frac{\kappa}{2} \int_{y^-}^\infty \frac{\d x^-}{p_+ } \,\Big[\epsilon^{(-) \, a}\Big( p_a + \frac{p_+}{k_+} k_a\Big)\Big]^2 e^{- \im \omega  C(x^-)}\,  e^{\epsilon(x^- - y^-)/2p_+}
    \times \mathcal{M}_n^y(l, \ldots).
\end{equation}
This has the same scaling as in the soft negative helicity photon, and is thus suppressed in the $\omega \rightarrow 0$ limit. The soft negative helicity graviton theorem on a self-dual gravitational plane wave is therefore 
\begin{tcolorbox}[colback=blue!5!white,colframe=blue!75!black,title=Soft negative helicity graviton theorem on a self-dual gravitational plane wave]\vspace{-0.35cm}
\begin{gather}
    \lim_{\omega \rightarrow 0} \omega \mathcal{M}_{n+1}(\{p_i\}; \omega \hat{k}^\mrin) = \frac{\kappa}{2} \sum_{p_i \text{ ingoing}} \frac{(\epsilon^{(-)} \cdot p_i)^2}{\hat{k} \cdot p_i}\, \mathcal{M}_n(\{p_i\}), \\
    \lim_{\omega \rightarrow 0} \omega \mathcal{M}_{n+1} (\{p\}, \omega \hat{k}^\mrout) = \frac{\kappa}{2}\sum_{p_i \text{ outgoing}} \frac{(\epsilon^{(-)} \cdot p_i)^2}{\hat{k} \cdot p_i} \, \mathcal{M}_n(\{p_i\}).
\end{gather}
\end{tcolorbox}
Similarly to the gluon case, this is compatible with the MHV formula for graviton scattering on a self-dual plane wave background~\cite{Adamo:2020syc,Adamo:2022mev} by the virtue of the scattering amplitude with a single negative helicity graviton vanishing.

\subsection{Zero memory and the perturbative expansion} \label{sec:pert}

In this section we will look at the zero-memory and perturbative version of the results in the previous section. In the previous sections, we emphasised that the background was viewed as `strong', with a non-zero memory effect, and this played a key role in the derivation of our results. We will now see how having zero memory affects this and how to recover the flat space and single photon/graviton background results by perturbatively expanding in the background field.

\subsubsection{Yang-Mills} \label{sec:weakYM}

First, we will adapt the results in the previous section to the case where the gauge theory background has zero memory:
\begin{equation}
    \int_{-x_i^-}^{x_f^-} \dot{\mathsf{A}}_a \,\d x^- = 0. 
\end{equation}
Equivalently, this means that the ingoing and outgoing gauges \eqref{sec1:gaugeInOut} are equivalent. It also means that an ingoing soft gluon will also be soft in the outgoing region and there is no distinction between the ingoing and outgoing states (except for the sign of their energy). 

When considering the scattering amplitude with a soft gluon, one can see that any divergences in the sandwich region are still suppressed. In contrast to the finite memory case we now get contributions from both the ingoing and outgoing regions and therefore the soft gluon theorem on a gauge theory plane wave background with no memory is
\begin{equation}
    \lim_{\omega \rightarrow 0} \omega \mathcal{M}_{n+1}^\mathsf{a} (\{ p_i\}; \omega \hat{k}) = g_{\text{YM}}\Bigg[\sum_{p_i \text{ ingoing}} \frac{\epsilon \cdot p_i}{\hat{k} \cdot p_i} \, T^{\mathsf{a}}_{p_i} +\sum_{p_i \text{ outgoing}} \frac{\epsilon \cdot p_i}{\hat{k} \cdot p_i}\, (T^{\mathsf{a}}_{p_i})^{\ast}\Bigg] \, \mathcal{M}_n(\{p_i\}). \label{sec3:GluNoMem}
\end{equation}
As there is no distinction between ingoing and outgoing gluons we have neglected the in/out superscript. Also note that having \emph{no background} is a special case of this result, and reproduces the usual soft gluon theorem~\cite{He:2015zea} in flat space.

\medskip 

We now expand perturbatively in the background, treating it as a single photon. We write it as the Fourier mode 
\begin{equation}
    \mathsf{A}_a(x^-) =  \delta \,  \mathsf{a}_a e^{\im q_- x^-}
\end{equation}
where $\mathsf{a}_a$ is a constant, corresponding to the polarisation vector of the background and $\delta$ is a small expansion parameter. Extracting the linear-in-$\delta$ part of the background amplitude is then equivalent to the tree-level amplitude with an additional photon with momentum $k_{\mu} = k_- n_{\mu} \neq 0$ and polarisation vector $\epsilon_{\mu} = \mathsf{a}_a \delta^a_{\mu}$. We can also treat this background as effectively having zero memory since 
\begin{equation}
    \delta\,  \mathsf{a}_a \int_{- \infty}^{\infty} e^{\im \, q_- x^-} \, \d x^- =2 \pi \im \,\delta \, \mathsf{a}_a \, \delta(q_-)
\end{equation}
which is zero under our assumption that $q_- \neq 0$. Applying the no-memory result \eqref{sec3:GluNoMem} on this perturbative background to linear-in-$\delta$ we find 
\begin{equation}
    \lim_{\omega \rightarrow 0} \omega \mathcal{M}^{\mathsf{a}\, \delta^1}_{n+2} ({p_i,q_-}, \omega \hat{k}) = g_{\text{YM}}\Bigg[\sum_{p_i \text{ ingoing}} \frac{\epsilon \cdot p_i}{\hat{k} \cdot p_i} \, T^{\mathsf{a}}_{p_i} +\sum_{p_i \text{ outgoing}} \frac{\epsilon \cdot p_i}{\hat{k} \cdot p_i}\, (T^{\mathsf{a}}_{p_i})^{\ast}\Bigg] \, \mathcal{M}^{\delta^1}_{n+1}(\{p_i, q_-\})
\end{equation}
where $\mathcal{M}^{\delta^1}$ represents the amplitude at linear order in $\delta$. This is compatible with the usual flat space result because our choice of gauge for the soft gluon ensures that the soft term that would arise from coupling to the background photon vanishes:
\begin{equation}
    \frac{\, \epsilon \cdot n}{\hat{k} \cdot n} = 0.
\end{equation}

\subsubsection{Gravity} \label{sec:pertGR}
In gravity, we explicitly have two forms of memory encoded in the asymptotic behaviour of the vielbeins, e.g. recalling equation \eqref{sec1:Easymp} 
\begin{equation}
    E^{\mrin}_{i \, a} (x^-> x_f^-) = b^{\mrin}_{i \, a} + c_{i \, a} x^-,\quad  E^{\mrout}_{i \, a} (x^- < x_i^-) = b^{\mrout}_{i \, a} + c_{i \, a} x^-
\end{equation}
the matrix $b$ roughly encodes the displacement memory effect, whilst $c$ roughly encodes the velocity memory effect. Here, we consider the case of $c = 0$. This means that 
\begin{equation}
    \sigma_{ab} (x^- < x_i^-) = 0 = \sigma_{ab}(x^- > x_f^-)
\end{equation}
for both ingoing and outgoing gauges. It also means that an ingoing plane wave $e^{\im k \cdot x}$ solution will still look like a plane wave solution in the outregion and vice versa. Explicitly 
\begin{equation}
    \Phi^{\mrin}(x^- > x_f^-) = \sqrt{\det (b^\mrin)} \exp \, \im \Big[ k_+ x^+ + k_i b^{\mrin \, i}_a x^a + \frac{k_i k_j}{2k_+} b^{\mrin \, i}_a b^{\mrin \, j\, a} x^-\Big]
\end{equation}
for a massless momentum $k$. The action of $b^\mrin$ is therefore a linear transformation of the initial momentum, so that the outgoing transverse momentum is $b^{\mrin \, i}_a k^a$. When $b = \mathbbm{1}$ we have no overall change in the momentum. Having $c = 0$ also affects the scalings in the geometric quantities that was crucial to derive the soft theorem, shown in Table \ref{sec3:noMemTable}.
\begin{table}
    \centering
    \begin{tabular}{c || c | c | c}
        & $x^- \rightarrow -\infty$ & $x^- \rightarrow + \infty$ &  \\ \hline
        $E^{\mrin}_{i \, a}$ & $ 1$ & $ b$ &\\
        $E^{\mrin \, i}_a$ & $ 1$ & $b^{-1}$&\\
        $\sigma^{\mrin}_{ab}$ & 0 &$ 0$& \eqref{sec1:deform} \\
        $\gamma^{\mrin}_{ij}$ & $1$ & $b^2$ & \eqref{sec1:ERmet} \\
        $\gamma^{\mrin \, ij}$ & $ 1$ & $b^{-2} $ & \\
        $F^{\mrin\, ij}$ & $ x^-$ & $ b^{-2}x^-$ & \eqref{sec1:Fdef}\\
        $\det{E^{\mrin}}$ & 1 & $\det(b) $&
    \end{tabular}
    \caption{The asymptotic-in-$x^-$ behaviour of useful geometric quantities in a gravitational plane wave with $c = 0$, ignoring the tensor structure of the quantities. The behaviour for the outgoing gauge are the same, with $+ \infty$ and $-\infty$ swapped and with implicit in/out labels on $b$.} \label{sec3:noMemTable}
\end{table}

It is  a simple adaptation of the usual arguments to see that the soft photon theorem on a gravitational plane wave with $c = 0$ is 
\begin{tcolorbox}[colback=ForestGreen!5!white,colframe=ForestGreen!75!black,title={Soft photon theorem on a gravitational plane wave with $c=0$}]\vspace{-0.35cm}
\begin{multline}
    \lim_{\omega \rightarrow 0} \omega \mathcal{M}_{n+1}(\{p_i\}; \omega \hat{k}^\mrin) \\= e \, \Bigg[\sum_{p_i \text{ ingoing}} Q_i\frac{ \epsilon \cdot p_i}{\hat{k} \cdot p_i} + \sum_{p_i \text{ outgoing}} Q_i\frac{ \sqrt{\det(b^\mrin)} \,\epsilon_b \cdot p_i}{(b^\mrin \hat{k}) \cdot p_i}\Bigg] \, \mathcal{M}_n(\{p_i\}), 
\end{multline}
\begin{multline}
    \lim_{\omega \rightarrow 0} \omega \mathcal{M}_{n+1} (\{p\}, \omega \hat{k}^\mrout) \\= e \,\Bigg[ \sum_{p_i \text{ ingoing}} Q_i\frac{\sqrt{\det(b^\mrout)} \,\epsilon_b \cdot p_i}{(b^\mrout \hat{k}) \cdot p_i}+\sum_{ p_i \text{ outgoing}} Q_i\frac{ \epsilon \cdot p_i}{\hat{k} \cdot p_i} \Bigg] \, \mathcal{M}_n(\{p_i\})
\end{multline}
\end{tcolorbox}
where $\epsilon_b$ is the polarisation vector for a photon with momentum $bk$. When $b = \mathbbm{1}$ this is the same as the flat space soft photon theorem, and this is also equivalent to the special case of a flat background. 

\medskip 

For the soft graviton case, the tail term and associated large coordinate transformation in the soft graviton theorem will persist but the rest follows analogously to the soft photon detailed above. This means that the soft graviton theorem on a gravitational plane wave with $c= 0$ is 
\begin{tcolorbox}[colback=ForestGreen!5!white,colframe=ForestGreen!75!black,title={Soft graviton theorem on a gravitational plane wave with $c=0$}]\vspace{-0.35cm}
    \begin{multline}
    \lim_{\omega \rightarrow 0} \omega \mathcal{M}_{n+1}(\{p_i\}; \omega \hat{k}^\mrin) = \frac{\kappa}{2} \, \Bigg[\sum_{p_i \text{ ingoing}} \frac{( \epsilon \cdot p_i)^2}{\hat{k} \cdot p_i} + \sum_{p_i \text{ outgoing}} \frac{ (\sqrt{\det(b^\mrin)} \,\epsilon_b \cdot p_i)^2}{(b^\mrin \hat{k}) \cdot p_i}\\ 
    + \sum_{p_i \text{ outgoing}} \frac{p_{i \, +}}{\hat{k}_+}\,  \int_{x_i^-}^{x_f^-} \frac{\epsilon^a \epsilon^b\sigma^{\mrin}_{ab}(s)}{|E^{\mrin}(s)|^{1/2}}\, \d s\Bigg] \, \mathcal{M}_n(\{p_i\}),
\end{multline}
\begin{multline}
    \lim_{\omega \rightarrow 0} \omega \mathcal{M}_{n+1} (\{p\}, \omega \hat{k}^\mrout) = \frac{\kappa}{2} \,\Bigg[ \sum_{p_i \text{ ingoing}} \frac{(\sqrt{\det(b^\mrout)} \,\epsilon_b \cdot p_i)^2}{(b^\mrout \hat{k}) \cdot p_i}+\sum_{ p_i \text{ outgoing}}\frac{ (\epsilon \cdot p_i)^2}{\hat{k} \cdot p_i} \\ 
    + \sum_{p_i \text{ ingoing}} \frac{p_{i \, +}}{\hat{k}_+}\,  \int_{x_i^-}^{x_f^-} \frac{\epsilon^a \epsilon^b\sigma^{\mrout}_{ab}(s)}{|E^{\mrout}(s)|^{1/2}}\, \d s\Bigg] \, \mathcal{M}_n(\{p_i\}).
\end{multline}
\end{tcolorbox}

Note that the last term in each of these respective formulae is integrated over the non-trivial region of the wave (where $\sigma_{ab} \neq 0$).

\medskip

We now focus on treating the background perturbatively as a single graviton. This is best done by writing the Einstein-Rosen metric with a Fourier mode perturbation
\begin{equation}
    \gamma_{ij}(x^-) = \delta_{ij} + \kappa \, \mathsf{a}_{i}\mathsf{a}_j \,e^{\im q_- x^-}
\end{equation}
corresponding to an additional graviton with momentum $k_\mu = k_- n_\mu \neq 0$ and symmetric polarisation tensor $\epsilon_{\mu \nu} = \mathsf{a}_{i} \mathsf{a}_j \delta^{i}_\mu \delta^j_\nu$ satisfying $\mathsf{a}_i \mathsf{a}^i = 0$. Up to first order in $\kappa$, this corresponds to 
\begin{gather}
    E_{i\, a}(x^-) = \delta_{i a} + \frac{\kappa}{2} \,\mathsf{a}_i \mathsf{a}_j \, e^{\im q_- x^-}, \qquad H_{ab}(x^-) = - \frac{\kappa}{2 q_-^2} \, \mathsf{a}_a \mathsf{a}_b  \,e^{\im q_- x^-}, \\
    E^i_a(x^-) = \delta^i_a - \frac{\kappa}{2} \, \mathsf{a}_i \mathsf{a}_j \, e^{\im q_- x^-}, \qquad \sigma_{ab} = - \frac{\kappa q_-}{2} \, \mathsf{a}_a \mathsf{a}_b \, e^{\im q_- x^-}, \qquad \det(E) = 1.
\end{gather}
Note that there is not a distinction in ingoing and outgoing coordinates in this case. Also, to leading order Brinkmann and Einstein-Rosen indices on $\mathsf{a}$ are equivalent: $\mathsf{a}_a = \delta_a^i \mathsf{a}_i$. 

\medskip 

First let's consider a soft photon in this background. As before, soft divergences occur when coupling to an external particle. These terms, for example attaching to an outgoing scalar, will look like \eqref{sec2:photonLast}. For the perturbative vielbeins above this expression simplifies to
\begin{equation}
    \omega\, eQ  \int_{y^-}^{\infty} \frac{\d x^-}{p_+} \epsilon^a E_a^j \Big( p_j + \frac{p_+}{k_+} k_i \Big) e^{- \im \omega C(x^-)} e^{\epsilon (x^- - y^-)/2p_+} \times \mathcal{M}_n^y(l, \ldots),
\end{equation}
where 
\begin{equation}
    C(x^-) = \Bigg(\frac{p_i p_j}{2p_+} \hat{k}_+ - \frac{p_i k_j}{p_+} + \frac{\hat{k}_i \hat{k}_j}{2 \hat{k}_+}  \Bigg) F^{ij}(x^-).
\end{equation}
Attaching to an ingoing scalar instead will involve an integral over the interval $(-\infty, y^-]$. Viewed as a series in the weak background, it can be argued that expanding the vielbeins $E$ or the integrated metric $F$ will \emph{not} contribute. This is because expanding these beyond leading order will invariably introduce a phase $e^{\im q_- x^-}$ into the integral. The integral at this order of the expansion is then proportional to 
\begin{equation}
    \int_{y^-}^{\infty} \d x^- \, e^{\im (q_- x^- - \omega C(x^-))} \, e^{\epsilon(x^- - y^-)/2p_+} \times \mathcal{M}_n^y(l, \ldots)
\end{equation}
which does not have a soft singularity in the limit $\omega \rightarrow 0$ since $q_- \neq 0$ so the exponent is finite. Therefore, we need only expand the $n$-point amplitude $\mathcal{M}_n^y (l, \ldots)$ in the background. 

The calculation continues straightforwardly and we obtain the soft photon theorem on this perturbative gravitational background
\begin{equation}
    \lim_{\omega \rightarrow 0} \omega \mathcal{M}^{\mathsf{a} }_{n+2} ({p_i,q_-}, \omega \hat{k}) = e\, \Bigg[\sum_{p_i \text{ ingoing}} Q_i \, \frac{\epsilon \cdot p_i}{\hat{k} \cdot p_i} +\sum_{p_i \text{ outgoing}} Q_i \, \frac{\epsilon \cdot p_i}{\hat{k} \cdot p_i}\Bigg] \, \mathcal{M}_{n+1}(\{p_i, q_-\}). 
\end{equation}
This is precisely the soft photon theorem in flat space, with $n$ charged scalar particles and one (uncharged) graviton.

\medskip

We now look at how a perturbative background affects the soft graviton theorem. Inspecting the tail terms from the zero-memory case, for a weak background, they are proportional to 
\begin{equation}
    - \frac{\kappa \, q_-}{4 \hat{k}_+} \epsilon_a \epsilon_b \mathsf{a}^a \mathsf{a}^b \int_{-\infty}^\infty \,  e^{\im q_- s} \, \d s = - \frac{2 \pi \im \, \kappa \, q_-}{4 \hat{k}_+} \epsilon_a \epsilon_b \mathsf{a}^a \mathsf{a}^b \, \delta(q_-) = 0.
\end{equation}
 Therefore these tail terms don't contribute and we  have (adapting the above argument for soft photons)
\begin{equation}
    \lim_{\omega \rightarrow 0} \omega \mathcal{M}^{\mathsf{a}}_{n+2} ({p_i,q_-}, \omega \hat{k}) = \frac{\kappa}{2} \, \Bigg[\sum_{p_i \text{ ingoing}} \frac{(\epsilon \cdot p_i)^2}{\hat{k} \cdot p_i} +\sum_{p_i \text{ outgoing}} \frac{(\epsilon \cdot p_i)^2}{\hat{k} \cdot p_i}\Bigg] \, \mathcal{M}_{n+1}(\{p_i, q_-\}). 
\end{equation}
This is compatible with the flat space soft graviton theorem since the coupling to the background vanishes as a consequence of the gauge $n \cdot \epsilon = 0$, i.e. the soft term from the background vanishes:
\begin{equation}
    \frac{(\epsilon \cdot n)^2}{\hat{k} \cdot n} = 0.
\end{equation}

\section{Discussion} \label{sec:Conc}

In this work we have obtained all-multiplicity results for scattering amplitudes on a non-chiral plane wave background in gauge theory and in gravity. These were derived from the Feynman rules of the theory, with arguments following Weinberg~\cite{Weinberg:1965nx} and are distinct from the flat space soft theorems through explicit dependence on whether the soft particle is ingoing or outgoing. In some sense, these results pose more questions than they answer. In this discussion, we will comment on some of them. 

\paragraph{Exponentiation and loop cancellations.}
The types of infrared singularities discussed in this paper are generally considered unphysical in flat space and can be cured with loop corrections or by dressing the external states. The cancellation using loops in QED at low points was discussed in  ~\cite{Dinu:2012tj,Ilderton:2012qe}. It would be interesting to see what the effect of dressing the external states in a background field is. This may follow some of the calculation on soft dressings in~\cite{Cristofoli:2025esy}. Whilst loop calculations on generic plane waves in Yang-Mills and gravity are quite prohibitively difficult (though an example calculation is seen in~\cite{Adamo:2019zmk}), it may be possible to find universal infrared behaviour from these types of contributions by applying similar methods to the ones used in this paper. 

\paragraph{The double copy.} 
The double copy~\cite{Adamo:2022dcm} and soft factorisation have long played a joint role in constraining all multiplicity expressions of graviton amplitudes from gluon amplitudes~\cite{Berends:1988zp}. However, the fate and manifestation of the double copy on strong backgrounds is still up for debate (see e.g. \cite{Ilderton:2024oly,Beetar:2024ptv,Borsten:2023paw,Lipstein:2023pih} for possible approaches). Examples at 3- and 4-points have been studied on plane waves in~\cite{Adamo:2017nia,Adamo:2018mpq,Adamo:2020qru}, whilst~\cite{Adamo:2024hme} presented a version of the double copy for all-multiplicity amplitude formulae in self-dual radiative backgrounds. The current state of things is hardly conclusive, but the soft theorems derived here may provide a hint as to what to expect. 

Firstly, the double copy on a plane wave background is often stated at the level of integrands. As we have seen in this paper, soft divergences (or lack thereof) arise by studying the soft limits of the whole integral. Therefore we are not really comparing the same thing and there is no tension between e.g. the soft graviton theorem having additional terms compared to the soft gluon theorem. Something that may be interesting to explore further is what happens to the soft gluon theorem on a \emph{non-Abelian} background~\cite{Coleman:1977ps,Trautman:1980bj}. For example in the flat space double copy, the colour-ordering of the external gluons (which are non-abelian) plays a crucial role in constructing graviton amplitudes. However, at this point the complexity of calculating these gluon amplitudes becomes equal (if not more) than calculating graviton amplitudes and the value of the endeavour may be purely conceptual.

Secondly, it's notable that a double copy structure persists for the negative helicity soft gluon and soft graviton on \emph{self-dual} backgrounds. This suggest the possibility that these perturbations away from the self-dual sector may have a persistent double copy structure even after integration. It would be interesting to explore this further, building on~\cite{Adamo:2024hme}.

\paragraph{Asymptotic symmetries.} In flat space, the leading soft theorems are the Ward identities of large gauge transformations at $\mathcal{I}^\pm$(see~\cite{Strominger:2017zoo,Miller:2021hty} for pedagogical reviews). The same should be the case for the soft theorems on plane wave backgrounds derived in this paper. The change in the soft theorems due to a background with memory should also be reflected in the derivation from large gauge symmetries. It would be interesting to see how these soft theorems play a role in an infrared triangle adapted to plane wave backgrounds.

One cause of concern may be that several examples of the soft theorems are consistently `half' of the flat space version (soft terms only existing when the scalar has matched asymptotics with the soft particle). However, in the derivation from the Ward identity in terms of soft and hard charges, both an ingoing and outgoing soft charge is inserted into the S-matrix, related through antipodal matching. A construction particularly suited to curved backgrounds can be found in~\cite{Kim:2023qbl}. In the soft theorems derived on gravitational backgrounds with no velocity memory, it seems that all that is needed to connect these ideas is a new form of antipodal matching dependent on $b$, the displacement memory. In this case, the tail terms in the soft graviton theorem would also presumable cancel upon relating $\sigma^\mrin$ and $\sigma^{\mrout}$. 

\paragraph{Self-duality.} Finally, it is worth discussing what these results tell us about specifically scattering on self-dual backgrounds. There has been much recent development in this area, inspired by twistor constructions and celestial holography~\cite{Bittleston:2024efo,Costello:2023hmi,Bogna:2024gnt,Bittleston:2023bzp,Bogna:2023bbd,Adamo:2025vzv}. In this paper we have shown that negative helicity soft theorems on self-dual backgrounds depend on the asymptotics of the soft particle (whether they are ingoing or outgoing). We have also demonstrated that there is no positive helicity soft gluon theorem in self-dual backgrounds. This indicates that perturbations away from self-duality are relatively simple but need to somehow account for the particles' asymptotic conditions. An obvious next step is seeing whether this is consistent with the N$^k$MHV amplitudes on self-dual plane waves presented in~\cite{Adamo:2020yzi,Adamo:2022mev}. It would also be interesting to repeat this analysis for self-dual backgrounds with sources, for example the self-dual dyon~\cite{Adamo:2024xpc}, self-dual Taub-NUT~\cite{Adamo:2023fbj} or flying focus fields~\cite{Adamo:2025vzv}.

\acknowledgments
I am particularly grateful to Tim Adamo for many discussions and comments on the draft of this paper. I also owe great thanks to Anton Ilderton for valuable conversations. I thank Andrea Cristofoli, Himali Dabhi, Joel Karlsson, and Donal O'Connell for insights on topics appearing in this paper. This work is supported by the Simons Collaboration on Celestial Holography CH-00001550-1.

\appendix
\section{Feynman rules on plane wave backgrounds} \label{app:Feyn}

\def\propog{\tikz[baseline=-0.5ex,mycirc/.style={circle,fill=black,minimum size=3pt, inner sep=0pt}]{
\node[mycirc, label=left:{$x$}] (A) at (0,0) {};   
\node[mycirc, label=right:{$y$}] (B) at (2, 0) {}; 
\node (C) at (1, 0.3)  {$l$} ;
\node (D) at (1.2, 0) {};
\node (F) at (0.5, 0) {};
\draw[->] (A)--(D);
\draw (F)--(B)}
}

In this appendix we review the relevant Feynman rules for calculating soft singularities on a plane wave background. We will only review the cases of a charged scalar and gluon in (scalar-)Yang-Mills, and a massive scalar coupled to gravity. The Feynman rules for photons are the abelian version of the gluon ones. They can be derived from the AFS prescription~\cite{Arefeva:1974jv,Jevicki:1987ax,Kim:2023qbl} for scattering amplitudes in terms of asymptotic states, applied to plane waves. The expressions for propagators are based on~\cite{Adamo:2018mpq}.

In both theories, the vertices of the Feynman diagram correspond to spacetime points $x_i$ that are integrated over. Additionally, these vertices have fields associated with them. In other words, each vertex $v_i$ with $N_i$ ingoing/outgoing fields with momentum $k_j$ comes with a factor 
\begin{equation}
    \int \d x_i^4\, \prod_{j = 1}^{N_i} \Phi^{(\epsilon_j)}_{k_j} (x_i) \label{app:vertex1}
\end{equation}
where $\epsilon_j$ labels whether the field is ingoing or outgoing (in the sense of the boundary conditions of the fields, e.g. \eqref{sec1:gaugeInOut} and \eqref{sec1:gInOut}). These fields need not be scalars --- they can also be photons, gluons or gravitons. In this case we would also add the appropriate polarisation tensors and indices to this expression, as we will see for the three-point amplitudes later. Note that only the $x^+$ component of the momentum is conserved at each vertex. Depending on the boundary conditions (if they are all ingoing, or all outgoing), the transverse $x^{\perp}$ components may also be conserved, but generically the $x^-$ component never is. In effect, for a Feynman diagram with $v$ vertices, one should always expect at least $v$ remaining integrals in $\{x_i^-: i \in v\}$.

The structure of the scalar propagators is also similar between the two theories. First we need to define the notion of off-shell momentum eigenstates in these theories. For an off-shell momentum $l$, where $l^2 = 2l_+ l_- - l_{\perp}l^{\perp} \neq m^2$, the off-shell charged scalar solution $\Phi^{\text{o.s.}}(x) = e^{\im \phi_l^{\text{o.s.}}(x)}$ in Yang-Mills is given by adding an extra term to the phase \eqref{sec1:phisol}:
\begin{equation}
    \phi^{\text{o.s.}}_l (x)  =    l_+ x^+ + (k_\perp + e \mathsf{A}_{\perp}) \, x^{\perp} + \frac{1}{2l_+} \int^{x^-} \d s \, (l_\perp + e \mathsf{A}_{\perp} (s))(l^\perp + e \mathsf{A}^{\perp} (s))  + \frac{l^2}{2l_+} x^-.
\end{equation}
In gravity, the situation is similar with the off-shell scalar solution $\Phi^{\text{o.s.}}(x) = \det(E)^{-1/2} e^{\im \phi_l^{\text{o.s.}}(x)}$ where 
\begin{equation}
    \phi_l^{\text{o.s.}}(x) = l_+ x^+ + l_i E^i_{\, a}(x^-) x^a + \frac{l_+}{2} \sigma_{ab}(x^-) x^a x^b + \frac{l_i l_j}{2l_+} F^{ij}(x^-)  + \frac{l^2}{2l_+} x^-.
\end{equation}
Each propagator then comes with an integral over the off-shell momentum $l$, with the Feynman prescription
\begin{equation}
     \propog \qquad \sim \qquad\im  \int \frac{\hat{\d}^4 l}{l^2 - m^2+ \im \epsilon}.\label{app:propRule}
\end{equation}

Taking just the $l^2$ dependent parts of the integral we have via the Feynman contour prescription
\begin{multline}
    \frac{\im}{2l_+} \int \frac{\hat{\d} (l^2)}{l^2 - m^2  + \im \epsilon} \exp \Big(\im \frac{l^2- m^2}{2l_+} (x^- - y^-) \Big)\\ =  \frac{1}{2l_+} \Theta \Big( \frac{y^- - x^-}{2l_+} \Big) \exp \Bigg[ \frac{\epsilon \, (x^- - y^-)}{2l_+}\Bigg] \label{app:propInt}
\end{multline}
as in flat space. For higher spin propagators, we would also  include the tensor structure and gauge-fixing ghost fields alongside the fields~\cite{Adamo:2018mpq}.

\medskip 

We will now consider the scalar-scalar-mediator vertex in our theories of interest. This will describe the tensor and functional structure that comes at each vertex in addition to \eqref{app:vertex1}. Both of them arise from the trilinear part of the action.

\paragraph{Yang-Mills.}

In Yang-Mills the structure of the 3-point with all momentum ingoing and an external gluon in lightfront and Lorenz gauge is 
\vspace{0.3cm}
\begin{equation}
    \begin{tikzpicture}[baseline=(current bounding box.center)]
        \node {
        \bd{2scalar}(60,60)
    \fmfset{arrow_len}{2.5mm}
    \fmfleft{i}
    \fmfright{o}
    \fmftop{t}
    \fmf{fermion}{i,g}
    \fmf{fermion}{o,g}
    \fmf{gluon}{g,t}
    \fmfv{lab=$1$}{i}
    \fmfv{lab=$2$}{o}
    \fmfv{lab=$\mu;\mathsf{a}$}{t}
    \fmfdot{g}
    \fmfv{l=$x_i$, l.a=-90, l.d=.1w}{g}
    \ed};
    \path[use as bounding box] ([shift={(2.5ex,2.5ex)}]current bounding box.north east) rectangle ([shift={(0,0)}]current bounding box.south west);
    \end{tikzpicture} 
     ~~~~~~~~~~~  \sim   - \im   g_{\text{YM}} \left( K_{1 \, \mu}(x_i) - K_{2\, \mu}(x_i) \right) T^{\mathsf{a}}   
    \end{equation} 
Note that the dressed momenta will depend on the asymptotic conditions of the fields involved. The dependence on the background field is entirely absorbed into the dressed momenta defined in \eqref{sec1:dMom}.

\paragraph{Gravity.}
The three-point vertex in gravity assuming that the external graviton is on-shell and satisfying the lightfront and de Donder gauge conditions is
\vspace{0.3cm}
\begin{equation}
    \begin{tikzpicture}[baseline=(current bounding box.center)]
        \node {
        \bd{2scalargrav}(60,60)
    \fmfset{arrow_len}{2.5mm}
    \fmfleft{i}
    \fmfright{o}
    \fmftop{t}
    \fmf{fermion}{i,g}
    \fmf{fermion}{o,g}
    \fmf{dbl_curly}{g,t}
    \fmfv{lab=$1$}{i}
    \fmfv{lab=$2$}{o}
    \fmfv{lab=$\mu\nu$}{t}
    \fmfdot{g}
    \fmfv{l=$x_i$, l.a=-90, l.d=.1w}{g}
    \ed};
    \path[use as bounding box] ([shift={(2.5ex,2.5ex)}]current bounding box.north east) rectangle ([shift={(0,0)}]current bounding box.south west);
    \end{tikzpicture} ~~~~~~~~~~~  \sim     
    - \im \frac{\kappa }{2}   \, (K_{1\, \mu}(x_i) - K_{2\, \mu}(x_i))\, (K_{1 \, \nu}(x_i) - K_{2 \, \nu}(x_i)). 
    \end{equation}

\paragraph{External particles.}

External particles with polarisation vectors  are attached to vertices by contracting the vertex tensor structure with the appropriate dressed polarisation vector: $\mathcal{E}_{\mu}^{(\epsilon)}(x)$ for gluons and $\mathcal{E}_{\mu \nu}^{(\epsilon)}(X)$ for gravitons. The $\epsilon$ superscript labels whether the particle is ingoing or outgoing, matching the field insertion at the vertex.

\section{Results for the convergence of integrals} \label{sec:conv}

In this appendix we collect some results on the asymptotic behaviour and convergence of integrals of interest in this paper.

\subsection{Slowly oscillating phases} \label{app:func1}

Consider the $\omega \rightarrow 0$ limit of an integral of the form
\begin{equation}
    I = \omega \int_{- \infty}^{y^-} \d x^- \, f(x^-) \, \exp [\im \,\omega g(x^-)] \times \exp \Bigg[\frac{\epsilon\, (x^- - y^-)}{2p_+}\Bigg]
\end{equation}
where $g(x^- < x_i^-) = A x^-$ and $f(x^- < x_i^-) = B$ for $A, B$ some constants. Here $p_+, \epsilon> 0$ and we implicitly take the $\epsilon \rightarrow 0$ limit before $\omega \rightarrow 0$.  

The integral can be split into two parts
\begin{multline}
   I = \omega \int_{x_i^-}^{y^-} \d x^- \, f(x^-) \, \exp [ \im \,\omega g(x^-)] \times \exp \Bigg[ \frac{\epsilon \, (x^- - y^-)}{2p_+}\Bigg] \\
    + \omega B \int_{- \infty}^{x_i^-} \d x^- \, \exp [\im \,\omega A x^- ] \times \exp \Bigg[ \frac{\epsilon(x^- - y^-)}{2p_+}\Bigg].
\end{multline}
In the limit as $\omega \rightarrow 0$, the first term is an integral over a compact interval therefore we can take the limit inside the integral smoothly and assuming that 
\begin{equation}
    \int_{x_i^-}^{y^-} f(x^-) \, \d x^-
\end{equation} 
exists and is finite, this term vanishes when $ \omega \rightarrow 0$. This is a reasonable assumption for our integrals of interest where $f$ has no singularities. For the second term, we cannot exchange the $\omega \rightarrow 0$ and the integration limits smoothly, but we can evaluate the integral explicitly first. Therefore 
\begin{align}
    \lim_{\omega \rightarrow 0} \lim_{\epsilon \rightarrow 0} I & = \lim_{\omega \rightarrow 0} \omega \lim_{\epsilon \rightarrow 0}\Bigg[ \frac{2p_+ B}{2\im p_+ A \omega + \epsilon} \exp \Bigg( \im \, A \omega x^- + \frac{\epsilon(x^- - y^-)}{2p_+}\Bigg)\Bigg]^{x_i^-}_{-\infty}\\
    & = \lim_{\omega \rightarrow 0} \omega \lim_{\epsilon \rightarrow 0}\frac{2p_+ B}{2\im p_+ A \omega + \epsilon} \exp \Bigg( \im \, A \omega x_i^- + \frac{\epsilon(x^- - y^-)}{2p_+}\Bigg)
\end{align}
where $\epsilon>0$ has allowed us to neglect the boundary term at $x^- = - \infty$. Now we take the $\epsilon\rightarrow 0$ limit (which corresponds to physically evaluating the amplitude). Then finally we have 
\begin{equation}
    \lim_{\omega \rightarrow 0} I = \frac{B}{\im A}.
\end{equation}

\subsection{More general slowly oscillating phases} \label{app:func2}

We consider the following limit:
\begin{equation}
    I =  \lim_{\omega \rightarrow 0} \lim_{R \rightarrow \infty} \omega \int_{y^-}^R f(x^-) e^{\im \omega g(x^-)} \d x^-.
\end{equation}
Here $f$ and $g$ are generic real functions, that may have ($\omega$-independent) poles or zeroes in a compact region. They also have asymptotic behaviour 
\begin{equation}
    \frac{f(x^-)}{g'(x^-)} \sim (x^{-})^{ - \delta}  \quad \text{as } x^- \rightarrow \infty
\end{equation}
where $\delta > 0$ and we also assume that $f(x^-)/g(x^-)$ has a definite sign for large enough $x^-$. Generally, assuming appropriate regularisation (such as contour deformation) we may restrict the integral of interest in the $\omega \rightarrow 0$ limit  over a section for some $L >  y^-$ that excludes all singularities in $f$ and $g$:
\begin{equation}
    I = \lim_{\omega \rightarrow 0} \lim_{R \rightarrow \infty} \omega \int_L^R f(x^-) e^{\im \omega g(x^-)} \d x^-. 
\end{equation}
We can make the dependence on $\omega$ simpler by rewriting  the integral as
\begin{equation}
    I = \lim_{\omega \rightarrow 0} \lim_{R \rightarrow \infty} \Bigg[ - \im \int_L^R \frac{f(x^-)}{g' (x^-)} \, \frac{\d}{\d x^-} \Big( e^{\im \omega g(x^-)}\Big)  \, \d x^-\Bigg]. 
\end{equation}
Now we integrate by parts, obtaining a boundary term and another integral
\begin{equation}
    I = \lim_{\omega \rightarrow 0} \lim_{R \rightarrow \infty} \Bigg[  - \im \int_L^R \frac{\d}{\d x^-} \Bigg( e^{\im \omega g(x^-)} \frac{f(x^-)}{g'(x^-)} \Bigg) \d x^- + \im \int_L^R e^{\im \omega g(x^-)} \frac{\d }{\d x^-} \Bigg(\frac{f(x^-)}{g'(x^-)} \Bigg)\Bigg]. \label{app2:IBPrel}
\end{equation}
We will now show that both parts separately converge in the limits. The second integral is absolutely convergent since the prefactor is a pure phase and the rest is an exact integral. This means that by the dominated convergence theorem~\cite{DCT}, we can exchange the $\omega \rightarrow 0$ and $R \rightarrow \infty$ limits so that
\begin{align}
     \lim_{\omega \rightarrow 0} \lim_{R \rightarrow \infty} \im \int_L^R e^{\im \omega g(x^-)} \frac{\d }{\d x^-} \Bigg(\frac{f(x^-)}{g'(x^-)} \Bigg) &=   \lim_{R \rightarrow \infty} \im \int^R_L \frac{\d}{\d x^-} \Bigg( \frac{f(x^-)}{g'(x^-)}\Bigg) \\
    & = - \im \frac{f(L)}{g'(L)}.
\end{align}
The upper boundary term was neglected since $f(x^-)/g'(x^-) \sim (x^-)^{-\delta}$ as $x^- \rightarrow \infty$ for $\delta >0$. 

The first term in \eqref{app2:IBPrel} is a boundary term, and again the limit at the upper boundary disappears:
\begin{equation}
    - \im \lim_{R \rightarrow \infty} \Bigg[ e^{\im \omega g(x^-)} \frac{f(x^-)}{g'(x^-)}\Bigg]^R_L = \im e^{\im \omega g(L)} \frac{f(L)}{g'(L)}.
\end{equation}
We can now straightforwardly take the $\omega \rightarrow 0$ limit of this, which means that  the total contribution from both parts of \eqref{app2:IBPrel} combined is
\begin{equation}
    I = \im  \frac{f(L)}{g'(L)} - \im \frac{f(L)}{g'(L)} = 0.
\end{equation}
\bibliographystyle{JHEP}
\bibliography{soft}
\end{document}